\documentclass[aps,prl,reprint,amsmath,amssymb,superscriptaddress,longbibliography,10pt,tightenlines]{revtex4-1}

\usepackage{latexsym}
\usepackage{epsfig}
\usepackage{braket}
\usepackage{graphicx,bbm,psfrag}
\usepackage{ragged2e}
\usepackage{caption}

\usepackage{subcaption} 

\usepackage{ragged2e}
\DeclareCaptionJustification{justified}{\justifying}
\usepackage{dsfont}
\usepackage{amsthm,booktabs,mathtools}
\usepackage{bm}
\usepackage{color}
\usepackage{comment}
\usepackage{dutchcal}
\usepackage{hyperref}
\hypersetup{
    colorlinks,
    citecolor=red,
    linkcolor=blue,
}
\DeclareMathOperator{\sgn}{sgn}

\newcommand{\bc}{\begin{center}}
\newcommand{\ec}{\end{center}}
\def\ba#1{\begin{array}{#1}\displaystyle}
\newcommand{\ea}{\end{array}}

\newcommand{\beq}{\begin{equation}}
\newcommand{\eeq}{\end{equation}}
\newcommand{\beqa}{\begin{eqnarray}}
\newcommand{\eeqa}{\end{eqnarray}}

\newcommand{\bi}{\begin{itemize}}
\newcommand{\ei}{\end{itemize}}

\def\frc#1#2{\frac{#1}{#2}}
\newcommand{\p}{\partial}

\newcommand{\br}{\langle}
\newcommand{\kt}{\rangle}

\newcommand{\ep}{\epsilon}

\newcommand{\dd}{{\rm d}}

\def\eqref#1{(\ref{#1})}

\begin{document}

\title{Observing quantum phase transitions at non-zero temperature:\\
non-analytic behavior of order-parameter correlation times}

\author{Istv\'an Cs\'ep\'anyi}
\affiliation{Department of Theoretical Physics, Institute of
Physics, Budapest University of Technology and
Economics, M{\H u}egyetem rkp. 3., H-1111 Budapest,
Hungary}
\affiliation{HUN-REN-BME-BCE Quantum Technology Research Group, Institute of Physics, Budapest University of Technology and Economics, M{\H u}egyetem rkp.~3., H-1111 Budapest, Hungary}

\author{Giuseppe Del Vecchio Del Vecchio}
\affiliation
{Laboratoire de Physique de l'Ecole Normale Sup\'erieure, CNRS, ENS and PSL Universit\'e, Sorbonne Universit\'e, Universit\'e Paris Cit\'e,
24 rue Lhomond, 75005 Paris, France
}

\author{Benjamin Doyon}
\affiliation
{Department of Mathematics, King's College London, Strand, London WC2R 2LS, UK
}

\author{M\'arton Kormos}
\affiliation{Department of Theoretical Physics, Institute of
Physics, Budapest University of Technology and
Economics, M{\H u}egyetem rkp. 3., H-1111 Budapest,
Hungary}
\affiliation{HUN-REN-BME-BCE Quantum Technology Research Group, Institute of Physics, Budapest University of Technology and Economics, M{\H u}egyetem rkp.~3., H-1111 Budapest, Hungary}

\begin{abstract} Phase transitions occur when a macroscopic number of local degrees of freedom coherently change their behavior. In ground states of quantum many-body systems, phase transitions due to quantum fluctuations are observed as non-analytic behaviors of order parameters, such as magnetization, as functions of a conjugate parameter, such as the magnetic field. However, as soon as thermal fluctuations are present, these effects are believed to disappear for local observables. We show that this is not necessarily the case: order parameters may still show non-analytic behaviors within their dynamics. With the example of the Ising model and using methods based on hydrodynamic fluctuations, we evaluate the exact order-parameter correlation time, in space-time directions of all velocities, in equilibrium states at nonzero temperature. We reveal non-analytic behaviors of spin correlation times as functions of the magnetic field, velocity, and temperature. As a function of the magnetic field, they occur at values that continuously approach that of the zero-temperature equilibrium transition point as the velocity is decreased and reach it within the light cone, where we obtain a new, temperature-independent logarithmic divergence characterizing the collective dynamics. Thus, collective effects induced by quantum fluctuations persist within the dynamics of local observables.
\end{abstract}

\maketitle

\paragraph{Introduction.---} The concept of phase transitions has played a crucial role in our understanding of emergent behavior in many-body physics. For instance, in equilibrium states, fluctuations, thermal or quantum, destroy order only if strong enough. But the passage from disorder to order is intricate: the many locally interacting degrees of freedom, such as spins on a lattice, start forming spatially organized regions where order prevails, which, however, still fluctuate on large scales. This multi-scale spatial organization leads to divergences in correlation lengths. These are second order phase transitions, observed as non-analytic behaviors of order parameters as functions of the temperature for thermal, or of some interaction strength for quantum phase transitions.

Importantly, in the conventional understanding, there is a sharp separation between thermal and quantum phase transitions: the latter may not occur at non-zero temperatures, because thermal fluctuations dominate. However, this is known to be specific to equilibrium quantities. In recent years, much progress has been made in understanding the non-equilibrium behavior of many-body systems, in particular, the dynamics of local observables. Dynamical phase transitions have been observed in various settings \cite{heyl_2013, Heyl_2018,Jafari2019,guo2019observation,tian2020observation}.
Yet, we believe that a fundamental understanding of their origin and form, paralleling that of equilibrium phase transitions, is still missing. A natural question is: Do large-scale collective behaviors observed at quantum critical points (QCPs) have residual dynamical effects at finite temperatures?

Within the so-called quantum critical fan, QCPs do have an effect \cite{sachdev_book}; however, those observed up to now disappear for large temperatures and dynamical quantities for large times. Highly non-local quantities related to quantum information (Loschmidt echo, quantum fidelity, quantum discord, etc.) do show QCP effects at finite temperatures \cite{mera2018, sedlmayr2018, nandi2024}, 
however, they are somewhat contrived observables from the viewpoint of many-body physics.

Perhaps the most natural quantity is the order parameter dynamical correlation function in an equilibrium state at temperature $T$. For instance, in a 1D lattice system,
\beq\label{eq:corr_func_intro}
    \br \sigma_0(0)\sigma_{j}(\zeta j) \kt
    \asymp e^{-j/\xi(\zeta,T)}\quad  (j\to\infty)\,,
\eeq
where $\sigma_j(t)$ is the local order-parameter at site $j$ and at time $t$. Here, $\xi(\zeta,T)$ is the temperature- and velocity-dependent correlation length for a correlation along a space-time ``ray'' of velocity $1/\zeta$, so that $\xi^{\rm time}(\zeta,T) = \zeta\,\xi(\zeta,T)$ is the correlation time, and in Eq.\ \eqref{eq:corr_func_intro} we ignore numerical and oscillatory factors as well as power-law corrections. At the QCP at zero temperature and on the spatial ray $\zeta=0$, criticality leads to a power-law instead of exponential decay \cite{sachdev_book}. But because of the mixing induced by temperature, and because order-parameter observables typically do not overlap with hydrodynamic modes (in the hydrodynamic-projection sense) \cite{Spohn1991LargeSD,hydro_projections}, an exponential decay as above, instead of a power-law \cite{doyon_2017_drude, Doyon_2018_large}, is often expected everywhere in space-time at non-zero temperatures even at a QCP.

In this paper, we propose that the non-analytic behavior of the non-equilibrium, dynamical quantity $\xi(\zeta)$  {\em may reveal effects of quantum critical physics at nonzero temperatures}. That is, the collective effects associated to quantum fluctuations at a QCP have a signature in this local dynamical quantity, even at large temperatures. We take the example of the paradigmatic transverse field Ising model (TFIM) \cite{LSM}, perhaps the most studied model of quantum phase transitions \cite{McCoy1983,LSM,sachdev_book,Musi}, with Hamiltonian
\begin{equation}\label{eq:H_ising_intro}
        H = -\frac{J}{2} \sum_{j = 1}^L \left( \sigma_j^1 \sigma_{j+1}^1 + h \sigma_j^3\right),
    \end{equation}
where $\sigma_j^k$ is the $k^{\rm th}$ Pauli matrix at site $j$, and the order-parameter observable is $\sigma_j = \sigma_j^1$. We set the coupling strength to $J=1$.
The model has a QCP at $h=1$, where the excitation spectrum is gapless, separating the ferromagnetic ordered phase ($h<1$) and the paramagnetic disordered phase ($h>1$).
We analyze the behavior of 
$
\xi(\zeta,T,h)$ 
as a function of the direction $\zeta$, the temperature $T$, and the magnetic field $h>0$.

\begin{figure}
    \includegraphics[width=0.95\linewidth]{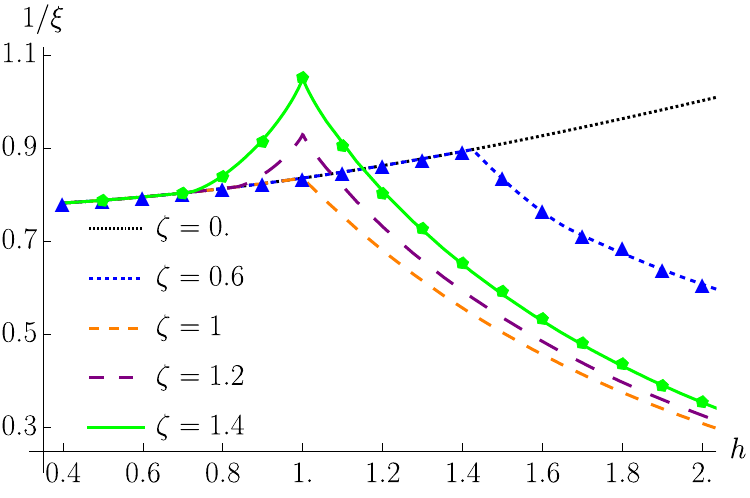}
    \caption{Inverse correlation length as a function of $h$ at temperature $T = 1$. The markers correspond to first-principle numerical results based on the Pfaffian technique of \cite{Derzhko1997}, while the solid lines are the predictions of Eq.\ \eqref{eq:factorization} with Eqs.\ \eqref{Omega} and \eqref{eq:fermionic_part}. For $h<1, \zeta>1$, the departures from the $\zeta=0$ black curve correspond to transitions between the space-like and time-like regions: the light cone in the Ising model is at $\zeta_* = \max (1, 1/h)$. For $\zeta>1$, the cusp at $h = 1$ represents the quantum phase transition,
    while for $\zeta\leq 1$, the corner at values $h\geq 1$ is a remnant of the quantum phase transition that brings the non-analyticity towards larger magnetic fields as we go farther in the space-like region.} 
    \label{fig:cusps_h}
\end{figure}

The equilibrium quantity, $\xi(\zeta=0,T,h)$, is known exactly \cite{Sachdev_1996} and does not show any non-analytic behavior. However, as $\zeta$ increases at fixed $T$, we observe (see Fig.~\ref{fig:cusps_h}) that 
a ``corner'' appears at a value $h=h_*(\zeta,T)$ 
that decreases from $h_*(0,T)=\infty$ to reach the equilibrium quantum phase transition point $h_*(1,T)=1$ on the model's light-cone $\zeta=1$. Increasing $\zeta$ further into the time-like region, it stays at $h_*(\zeta>1,T)=1$ but becomes a cusp where the left and right derivatives 
of the correlation time diverge. We show below that for any finite $T$, the divergence of the derivative of the inverse correlation time takes a temperature-independent form,
\beq\label{divtime}
    \frc{d}{dh}
    \Big(\frc1{\xi^\text{time}}\Big)
    \sim
    \frac{\sqrt{\zeta^2-1}}{\pi\zeta}
    \sgn(1-h)\log|h-1|\quad (h\to 1)
\eeq
(note how the coefficient stays finite in the purely time-like limit $\zeta\to\infty$). We explain below how despite the finite temperature, \eqref{divtime} is indeed a collective effect from the QCP. We observe from Fig.~\ref{fig:cusps_h} that at these corners and cusps, the correlation time reaches its minimum. The divergence \eqref{divtime} at the QCP within the light cone, and the way this cusp emerges via corners that appear at $h=\infty$ as we go beyond the purely spatial direction, are our main results.

\begin{figure}[t]
    \centering
    \includegraphics[width=0.95\linewidth]{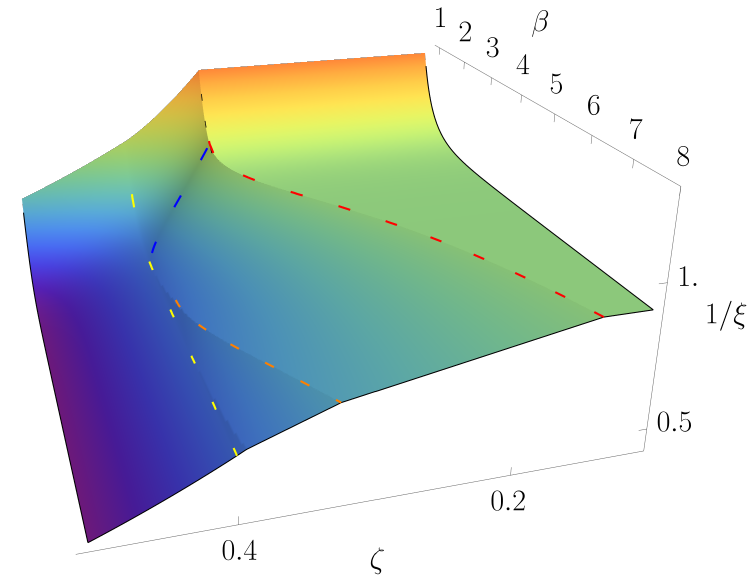}
    \caption{Inverse correlation length  $1/\xi(\zeta,T,h)$ determined analytically from Eq.\ \eqref{eq:factorization} as a function of the space-time direction $\zeta=t/x$ and the inverse temperature $\beta=1/T$ at magnetic field $h=2.5$. The dashed lines indicate nonanalytic  ``corners'' where the gradient is discontinuous.}
    \label{fig:3Dplot}
\end{figure}

We also show that for fixed $h>1$ (in the disordered phase), the inverse correlation length has infinitely many corners both as a function of $\zeta$ and $T$ (see Fig. \ref{fig:3Dplot}). A similar behavior has recently been uncovered in Ref. \cite{Csepanyi2024} in the Ising field theory that gives the scaling limit near the QCP. The present work, in particular, extends these results to the spin chain with generic magnetic fields.

\paragraph{A technical challenge.---}

A full understanding of order-parameter dynamical correlations in many-body systems has been hampered by a lack of strong methodologies. Perhaps surprisingly, even in the TFIM, accounting for its free-fermion description \cite{LSM}, the exact form of $\xi(\zeta,T,h)$ remained until now inaccessible.

In the Ising model, the crux of the problem is that the mapping between spins and fermions is nonlocal. Indeed, the Jordan–Wigner (JW) transformation \cite{JW}
$\sigma_j^3 = 1-2c_j^\dagger c_j,$
$\sigma_j^1 = S_{1,j-1}\left(c_j + c_j^\dagger\right)$
that maps spin operators to canonical fermionic operators $c_j,c^\dagger_j$
involves the nonlocal string operator 
$    S_{i,j} = \prod_{l = i}^{j} \left ( 1-2c_l^\dagger c_l \right )$.
Under this mapping, the Hamiltonian is transformed into quadratic forms of the fermionic operators, one for the even and one for the odd-fermion subspace. By Fourier transform and Bogoliubov rotation, both 
are brought to the canonical, diagonal form
\begin{equation}\label{Hdiag}
    H =  \sum_{p} \epsilon(p)\,\alpha_{p}^\dagger \alpha_{p}\,,
\end{equation}    
where the dispersion relation of the free spinless fermions is
 $   \epsilon(p) = \sqrt{1+h^2-2 h \cos p}.$
Rewriting the correlation function \eqref{eq:corr_func_intro} in terms of fermions yields nonlocal string operators evolved in time, and evaluating the resulting dynamical multi-point function, and in particular its asymptotics, using free fermion techniques becomes a hard mathematical problem. In fact, until recently, there was no technique to determine $\xi(\zeta,T,h)$ exactly in all regimes of the model.

Low-temperature semiclassical methods 
\cite{SachdevYoung1997, Buragohain_1999} can be applied, with the advantage that they provide an intuitive, if approximate, physical picture, but it is not clear whether the qualitative conclusions that can be reached hold at larger temperatures. Free fermion techniques give rise to 
differential equations for the correlator \cite{Perk1980,Perk1984,Perk}, Fredholm determinants \cite{doyon2007finitetemperature, chernowitz_2022, Zhuravlev2022,Granet2022} and other representations based on form factor series \cite{thermalFF,Gohmann_2019,Gamayun2021}
which sometimes allow for rigorous asymptotic analysis via Riemann--Hilbert problems \cite{Korepin}. However, these are extraordinarily intricate and do not easily account for the underlying physics. Perhaps one of the most thorough analyses of the TFIM to date, although still approximate, was given in Ref. \cite{Essler}, based on a partial resummation of the form factor series. Our results agree with these in most regimes; however, in some cases, they disagree. In particular, most of the nonanalytic features are not captured by Ref. \cite{Essler}.

\paragraph{Hydrodynamic breakthrough.---}

A breakthrough happened with the advent of hydrodynamic methods \cite{castro_alvaredo_2016, bertini_2016, doyon_prx_2025}, which give rise to {\em exact asymptotic expansions} bypassing the need for a full microscopic analysis of the model. In particular, the Ballistic Fluctuation Theory (BFT) \cite{Myers1} has been shown in \cite{Myers2} to connect order-parameter correlation times to the large-deviation theory of hydrodynamic fluctuations, thereby making exact expressions for $\xi(\zeta,T,h)$ accessible. Crucially, because the methods are based on hydrodynamics, {\em they fundamentally extract the correct emergent degrees of freedom for the many-body dynamics} -- the hydrodynamic modes -- and hence encode many-body collective effects at the basis of phase transitions.

Applying hydrodynamic methods, we explain below how $\xi(\zeta,T,h)$ has two contributions: the {\em fluctuation} and {\em propagation} part (these were previously referred to as ``classical'' and ``quantum'' \cite{SachdevYoung1997}, which is in fact a slightly misleading nomenclature). The fluctuation part is a dynamical free energy, related to the large deviation function for the total spin current. The propagation part is due to oscillatory hydrodynamic modes \cite{ampelogiannis2021ergodicity} propagating along the observation ray in a state whose Boltzmann weight is modified by the spin current \cite{XXmodel}. These modes represent the formation and absorption of quasi-particles. The behavior of $\xi(\zeta,T,h)$ is determined by the interplay between the fluctuation and propagation parts. Much like the equilibrium free energy, the dynamical free energy has non-analytic points where the gap closes at $h=1$, giving rise to the cusps: the sharp change in current fluctuations is due to QCP physics. Further, because there is a continuum of oscillatory modes in the Ising model -- the continuum of free-fermionic modes -- with velocities in 
$[-\mathrm{min}(1,h),\mathrm{min}(1,h)]$
, in the time-like region, the propagation part dissipates into a power-law decay and does not affect correlation times, thus the cusp remains intact. However, outside the light-cone $\zeta<1$, oscillatory modes give exponential contributions that, technically, depend on a subtle saddle-point analysis, with the effect of mollifying the cusp to a corner and making its position $h_*(\zeta)$ continuously vary from $h_*(1)=1$ to $h_*(0)=\infty$. We now explain how this works.

\paragraph{Ballistic Fluctuation Theory.---}
In the presence of conserved quantities, the algebraically decaying asymptotic form of dynamical correlation functions of observables under the ballistic scaling can be obtained with the method of hydrodynamic projections \cite{Spohn1991LargeSD,hydro_projections}, if the observable couples non-trivially with at least one of the hydrodynamic modes. Sometimes, e.g. for order parameters \cite{XXmodel}, this coupling is absent. Ballistic Fluctuation Theory (BFT) \cite{Myers1,Myers2} is a powerful theoretical framework arising from the combination of hydrodynamic projections and Large Deviation Theory (LDT) that can address this situation.

In the language of LDT, the main object of the BFT is the scaled cumulant generating function (SCGF) of the variable $\omega(x,t) = \int_{\gamma} \left[q(x',t')\dd x' - j(x',t') \dd t'\right]$ where $\gamma:(0,0)\to (x,t)$ is a path in space-time and $q(x,t), j(x,t)$ are a conserved density and its current, satisfying $\p_t q + \p_x j = 0$. If $\gamma$ is totally spatial, then $\omega(x,0)$ is the total charge within an interval of length $x$, for which the SCGF is the free energy; while if it is totally temporal, $\omega(0,t)$ is the integrated current (transported charge) in time $t$ through the origin, for which the SCGF can be interpreted as a dynamical free energy.
The central prediction of BFT is an exact expression for the SCGF of $\omega(x,t)$, valid for $x,t\to +\infty$ with $t/x =\zeta$ held fixed:
\begin{equation}\label{eq:scgf}
    \mathcal{F}(\lambda;\zeta) = \lim_{\ell\to\infty}\ell^{-1}\log\braket{e^{\lambda \omega(\ell x,\ell t)}}.
\end{equation}
In Eq.~\eqref{eq:scgf}, the average can be taken over any Gibbs or generalized Gibbs state; here we concentrate on thermal Gibbs states.
For the fermion number in free fermion models, the BFT prediction \cite{Myers1} generalizes to arbitrary space-time directions the celebrated Levitov--Lesovik formula (with pure transmission) \cite{levitov1993charge,
levitov_1996}. For our purposes, as we will see below, of interest is the value $\lambda=i\pi$ of the counting parameter. The BFT prediction for $\braket{e^{i\pi\omega(x,t)}}=\Omega(x,t)$ is
\begin{equation}\label{Omega}
\Omega(x,t) = \exp \left [ \int_{-\pi}^\pi \frac{dp}{2\pi}  |x - \epsilon'(p) t| \log \left | \tanh\frac{\beta \epsilon(p)}{2}\right |\right ]\,,
\end{equation}
where $\epsilon(p)$ is the single-particle dispersion relation of the free fermions. The BFT has been applied beyond free models, including transport of generic charges in hard rods and more general integrable models \cite{Myers1, Myers2} as well as in stochastic chains such as TASEP \cite{myersthesis}, and it gives predictions for correlations of twist fields \cite{Myers2,e27121230}, which has been applied to the two-point function of vertex operators in the sine--Gordon model \cite{del_vecchio2023}.

\paragraph{The propagation--fluctuation factorization.---}

Recently, the asymptotic behavior of dynamical spin-spin correlation functions in the XX model was studied \cite{XXmodel}. By the JW transformation, this model can also be mapped to a system of free fermions. It has a $U(1)$ conserved charge, the magnetization, which translates to the fermion number under this mapping.

A crucial proposal of \cite{XXmodel} is what we may term the {\em propagation--fluctuation factorization}: spin-spin correlation functions factorize into a special two-fermion correlator within a state with purely imaginary chemical potential $\beta\mu=i\pi$, and the SCGF $\Omega(x,t)$:
\begin{equation}\label{eq:XX_correlation}
\braket{\sigma^-_0\sigma^+_x(t)}\asymp
\braket{a_0(0)a^\dagger_x(t)}_{i\pi}\Omega(x,t)\,.
\end{equation}
The chemical potential $\beta\mu=i\pi$ is due to the effects of the time-evolved JW strings on the state, while the factor $\Omega(x,t)$ is the contribution of the JW strings themselves.
This factorization makes the connection between order-parameter correlations and the BFT. It is an important aspect of the general theory for order parameter.

\paragraph{The Doubling trick.---}

One of the main technical results of this paper is to clarify the propagation--fluctuation factorization and the use of the BFT in the TFIM, and to provide a full first-principle numerical check.
Unlike the XX model, the TFIM does not admit a conserved fermion number.
We can nevertheless make sense of formula \eqref{eq:XX_correlation}: the ``$i\pi$'' prescription for the SCGF counting parameter means the inclusion of the term $e^{i\pi\sum_p \alpha^\dagger_p\alpha_p}$ in the thermal trace which is nothing but the fermion parity, also the parity of up spins, $S_{1,L}=e^{i\pi \sum_{j=1}^L \delta_{\sigma^3_j,1}}=e^{i\pi \sum_{j=1}^L c_j^\dagger c_j}=e^{i\pi\sum_p \alpha^\dagger_p\alpha_p}$ and is conserved in the TFIM (see Eq.~\eqref{Hdiag}). The fact that  Eqs. \eqref{eq:XX_correlation} and \eqref{Omega} still hold in the TFIM can be shown
using a trick that involves the 
`doubling' of the system \cite{ItyksonZuber1977} by considering two independent copies (for details, see the End Matter and the SM). The idea is that the resulting model is the discretized version of the free Dirac theory that possesses a conserved $U(1)$ charge -- generating rotations among the copies -- and the direct product of string operators in the two copies is identified with $e^{i\omega(x,t)}$ for this charge, for which the BFT applies.
This leads to our first technical result:
\begin{equation}
    \begin{aligned}
        \braket{\sigma_j^1 \sigma_{k}^1(t)} &\asymp F_\text{d}(x,t) \Omega_\text{d}(x,t)\,,\;\quad\qquad\qquad\;
        h>1,\\
        \braket{\sigma_j^1 \sigma_{k}^1(t)} &\asymp F_\text{o}(x,t) \Omega_\text{o}(x,t) = \Omega_\text{d}(x,t)\,,\quad
        h<1,
    \end{aligned}
    \label{eq:factorization}
    \end{equation}
where $x = k-j$ and the ``o/d'' subscripts refer to the ordered (ferromagnetic) and disordered (paramagnetic) phases. Again, the correlator factorizes into a \emph{propagation} part $F_\text{o/d}(x,t)$ and a \emph{fluctuation} part $\Omega_\text{o/d}(x,t)$. In the disordered phase,
\begin{equation}
    F_\text{d}(x,t) =\int_{-\pi}^\pi \frac{dp}{2\pi} \left [ \frac{e^{i \theta_p} e^{ipx-i \epsilon(p)t}}{e^{\beta \epsilon(p)}-1} +\frac{e^{i \theta_{p}} e^{ipx+i \epsilon(p)t}}{1-e^{-\beta \epsilon(p)}}\right],
    \label{eq:fermionic_part}
\end{equation}
where $\theta_p$ is the Bogoliubov angle satisfying $\tan(\theta_p)=\sin p/(\cos p -h)$, and $\Omega_\text{d}(x,t)=\Omega(x,t)$ given in Eq. \eqref{Omega}.
In the ordered phase, $F_\text{o}(x,t) = e^{x \delta}$ and $\Omega_\text{o}(x,t) = e^{-x \delta} \Omega(x,t)$. Here $\delta$ describes the scaling of the gap: $\Delta \sim e^{-L \delta}$, but its contribution cancels in Eq.\ (\ref{eq:factorization}).

\paragraph{Correlation length.---} 

The decay of the finite temperature correlation function is always exponential thanks to the fluctuation factor $\Omega(x,t)$. We note that this factor has appeared previously in the literature. It was obtained from the partial resummation of a form factor series in Ref. \cite{Essler}, and approximate versions of it appeared earlier \cite{SachdevYoung1997,Altshuler_2006,Reyes}. At low temperature, it can be interpreted as the number of quasiparticle trajectories crossing the space-time interval between the two operators, and it is at the heart of the semiclassical analysis \cite{SachdevYoung1997}.

In the paramagnetic phase, the propagating factor $F_\text{d}(x,t)$ is also present. For time-like separations $\zeta>1$, it decays algebraically and does not modify the correlation length, which is therefore obtained from the fluctuation part $\Omega(x,t)$ for all values of $h$. An asymptotic analysis leads to the result in Eq.\ \eqref{divtime}, giving the cusps at $h=1$ for $\zeta>1$ (see Fig.\ \ref{fig:cusps_h}). The divergence in Eq.\ \eqref{divtime} comes the vanishing of $\ep(p)$ at $p=0,h=1$, indicating that in the dynamical free energy, the continuum of low-momenta states, representing coherent fluctuations on large distances, dominate. This divergence is enabled by the Bose-Einstein-like condensation that occurs thanks to the effective change of statistics that we can see by interpreting \eqref{Omega} in terms of free energies, and induced by the fermion counting variable $e^{i\pi\omega(x,t)}$.

For space-like separations $\zeta<1$, $F_\text{d}(x,t)$ decays exponentially, and the correlation length has two contributions: 
$\xi^{-1} = \xi_\text{pr}^{-1}+\xi_\text{fl}^{-1}$. Both terms contribute to the corners in $h$ shown in Fig.\ \ref{fig:cusps_h} for $h>1$ and $\zeta<1$. The fluctuation contribution $\xi_\text{fl}^{-1}$ also has a nonanalytic dependence on $\zeta$ as the light cone set by the maximal velocity $\max\epsilon'(p)=\min(1,h)$ is crossed. For example, at the QCP $h=1$ we find that for $\zeta < 1$, $\dd\xi_\text{fl}/\dd\zeta=0$, while for  $\zeta = 1+\varphi$, $\varphi\ll1$,
\begin{equation}
    \frac{\dd\xi_\text{fl}^{-1}(\zeta)}{\dd\zeta} \approx -\frac{
    \sqrt{2}}{\pi} ( \log \varphi+\log (2 \beta^2)-2 )\varphi^{1/2}\,,
    \label{eq::fluctuation-derivative-wrt-zeta}
\end{equation}
showing that the second derivative of the correlation length jumps.

Furthermore, the propagating contribution $\xi_\text{pr}(\zeta, h, T)$ has infinitely many nonanalytic points on the $(\zeta,\beta)$-plane. In order to extract $\xi_\text{pr}$, we have to analyze the asymptotic behavior of the integral \eqref{eq:fermionic_part} for $F_\text{p}(x,t)$ (see the SM for details). We observe that the two terms in Eq.\ (\ref{eq:fermionic_part}) are connected by an $\epsilon(p) \rightarrow -\epsilon(p)$ transformation. This means that $F(x) = \int_C \frac{1}{2\pi}\frac{e^{i\theta_z} e^{izx - i \epsilon(z) t}}{1-e^{\beta \epsilon(z)}} dz$
where $C$ consists of two disjoint complex contours that connect $-\pi $ and $\pi$ on the two sheets of the square root in the dispersion relation. For $\zeta > 1$, saddle points lie on the real axes, so we can use the stationary phase method in both sheets to conclude that the decay of $F(x)$ is algebraic. For $1<1/\zeta < h$, we have to modify the integration contour and use the method of steepest descent. This contour deformation does not cross any poles in the integrand of $F(x)$, so the decay solely comes from the saddle point contribution:
\begin{equation}
    \xi_\text{pr}^{-1} = -\sqrt{1-\zeta^2} + \text{arccosh}(1/\zeta).
\end{equation}
For $1<h<1/\zeta$, the contour deformation required crosses a certain number of poles on both sheets, and the resulting contour is closed on the two-level Riemann surface. There is a competition between the crossed poles and saddle points to give the leading decay.
As the parameters vary, we transition between domains where a certain pole or saddle point dominates, at whose boundaries the derivatives of $\xi_\text{pr}$ jump, resulting in corner-type nonanalyticities.
This is illustrated in Fig.\ \ref{fig:cusps_h} (in $h$) and Fig.\ \ref{fig:3Dplot} (in $\zeta,\beta$). 
Explicit expressions for the results
can be found in the End Matter and in the Supplementary Material.
We checked all the above analytical results by first-principles numerical calculations based on the free fermionic formulation \cite{Derzhko1997} (as indicated by the markers in Fig.\ \ref{fig:cusps_h}).
Further, a preliminary analysis shows that similar nonanalytic behaviors occur in the XX model.

\paragraph{Conclusions.---}

We have evaluated the exact correlation length/time of the order parameter in space-time in the TFIM and shown that it exhibits nonanalytic behaviors related to the QCP of the model even at finite temperatures. We have numerically established in this model the propagation-fluctuation factorization \eqref{eq:factorization}, which is at the basis of the results. The cusps observed within the light cone are consequences of the QCP and related to the fluctuations of the transverse spin current. This phenomenon is distinct from the non-analyticities found in the equal-time correlation length in other models \cite{Klumper2001,Sirker2002,Patu2014}.
We note that previous low-temperature approximations \cite{SachdevYoung1997,Altshuler_2006,Reyes} provided expressions for the fluctuation part, however, these cusps were not discussed; these approximations are uncontrolled as they exchange low-temperature and long-time limits. We have also uncovered, by a stationary-phase analysis of the exact propagation part, the manner in which cusp singularities enter the light cone, starting, at large velocities, as corners coming from infinite-magnetic field. Furthermore, corners are also present for the correlation length as a function of $\zeta$ and $T$.
Experimental verification of our results may be possible in the near future on quantum simulators based on neutral Rydberg atoms \cite{Schauss2018,Browaeys2020} and ancilla-based measurement methods \cite{Uhrich2017}.

Our results are based on hydrodynamic ideas, and thus extract the universal degrees of freedom at their basis: the spin current fluctuations and the propagation of quasi-particle oscillatory modes in a spin-current-modified state. The results of this paper, in addition to elucidating quantum phase transitions in dynamical correlation times for the much-studied Ising model, constitute a proof of concept for broader applications. We expect the above picture to generalize to interacting integrable models, as a continuum of quasi-particle oscillatory modes is expected to be present, and dynamical free energies are accessible by the thermodynamic Bethe Ansatz \cite{del_vecchio2023}. It would also be interesting to see whether these hydrodynamic ideas can be generalized to higher dimensions to access counting statistics obtained from random matrix theory \cite{Gouraud_2022, dixmerias2025}. Research in these directions is still ongoing.

{\em Acknowledgments.---} This work was supported by the HUN-REN Hungarian Research Network through the Supported Research Groups Programme, HUN-REN-BME-BCE Quantum Technology Research Group (TKCS-2024/34), and also by the National Research, Development and Innovation Office (NKFIH) through the OTKA Grant K 138606. 
I. Cs. was partially supported by the Doctoral Excellence Fellowship Programme (DCEP), funded by the National Research, Development and Innovation Fund of the Ministry of Culture and Innovation, and the Budapest University of Technology and Economics, under a grant agreement with the National Research, Development and Innovation Office.
BD was supported by UKRI Horizon Europe guarantee (ERC Advanved Grant Scheme) EP/Z534304/1. 
G.D.V.D.V. akcowledges the support of  ANR Grant No.\ ANR-23-CE30-0020-01 EDIPS.

\bibliography{biblio}

\onecolumngrid

\newpage
\begin{center}{\bf End Matter}\end{center}

\paragraph{The doubling trick.---}

Consider two independent copies of the model on a direct product Hilbert space with fermion operators
$c^{(i)}_j,{c^{(i)}_j}^\dagger$, $i=1,2$, 
and associated string operators $S^{(i)}_j$. 
A Dirac fermion can then be constructed as
\begin{equation}
    \Psi = \frac12\left(\begin{matrix}
    a^{(1)} + ia^{(2)} \\ \bar a^{(1)} +i\bar a^{(2)}
    \end{matrix}\right),
\end{equation}
where $a^{(i)}=c^{(i)}+{c^{(i)}}^\dagger$, $\bar a^{(i)}=i({c^{(i)}}^\dagger-c^{(i)})$ are Majorana operators.
In terms of $\Psi$, the Hamiltonian $H_\text{d}$ of the doubled model becomes the discretized version of the free Dirac Hamiltonian which has a conserved $U(1)$ charge $Q = \sum_{j = 1}^L \Psi_j^\dagger \Psi_j$.
We can construct the special twist fields
\begin{equation}
    \mathcal U_{j,k} = :e^{i\pi \sum_{l=j}^{k} \Psi_l^\dagger \Psi_l}:\,,
\label{eq:twist}    
\end{equation}
where the normal ordering with respect to $c^{(i)},{c^{(i)}}^\dagger$ simply gives rise to an overall sign. It can be shown (see SM) that this twist field is nothing but the product $S^{(1)}_{j,k}S^{(2)}_{j,k}$ of the string operators in the two copies. 

Here we show how we can obtain the equal-time correlator using the doubling trick, while the general non-equal time case is discussed in the SM.
Consider the square of the correlator we are interested in,
\begin{equation}
    \braket{\sigma_j^1 \sigma_k^1}^2 = \braket{(\sigma_j^1)_1(\sigma_j^1)_2(\sigma_k^1)_1(\sigma_k^1)_2}^\text{d} \,,
\label{eq:squared}
\end{equation}
where the subscripts ``1'', ``2'' refer to the two copies and the superscript ``d'' refers to the doubled model. 
When written in terms of fermions, a JW string appears in Eq. \eqref{eq:squared}.
Exploiting that their product is the twist field \eqref{eq:twist} in the doubled theory, we are in a position to generalize the techniques of Ref.\ \cite{XXmodel} developed for the XX model to the Dirac system and to a four-spin correlation function. Invoking the factorization property analogous to that in Eq.~\eqref{eq:XX_correlation} leads to 
\begin{equation}
    \braket{\sigma_j^1 \sigma_k^1}^2 = \braket{\bar{a}^{(1)}_j a^{(1)}_k\bar{a}^{(2)}_j a^{(2)}_k}^\text{d} _{i\pi}\braket{\mathcal{U}_{j,k}}^\text{d}\,,
\end{equation}
where $\braket{\dots}^\text{d}_{i\pi}=\mathcal{Z}^{-1}\mathrm{Tr} \left (e^{-\beta H_\text{d} + i\pi Q}\dots\right)$, i.e., it is a grand canonical average in the doubled model with imaginary chemical potential $\beta\mu=i\pi$. Using that (up to a sign) $e^{i\pi Q}=\mathcal{U}_{1,L}=S^{(1)}_{1,L}S^{(2)}_{1,L}$, the fermionic correlator factorizes, so 
\begin{equation}
    \braket{\sigma_j^1 \sigma_k^1}^2 
    = \braket{\bar{a}_j a_k S_{1,L}}^2 \braket{\mathcal{U}_{j,k}}^\text{d}\,.
\label{eq:factor}    
\end{equation}
The second factor can be evaluated using the BFT in the free Dirac theory, while the first factor is a fermionic correlator in the TFIM. Noticing that $S_{1,L}$ is simply the fermion parity, it can be written as $e^{i\pi\sum_p \alpha^\dagger_p\alpha_p}$, and its effect in the canonical average is thus formally equivalent to the introduction of an imaginary chemical potential $\beta\mu=i\pi$ for the fermion number. This leads to a factor of $e^{i\pi}$ in the Fermi--Dirac distribution and turns it into the Bose--Einstein distribution. Taking the square root yields the TFIM two-point function. In the SM, we show how the terms on the right-hand side result in Eq.\ \eqref{eq:factorization}.

\paragraph{Cusps and logaritmically divergent derivatives.---} Cusps in Fig.~\ref{fig:cusps_h}, that arise at the QCP $h=1$ in the time-like regime $\zeta>1$, are due to the fluctuation contribution $\Omega(x,t,\beta,h)$, Eq.~\eqref{Omega}, and in particular to the closing of the gap. Let us show how this works. The correlation time $\xi^\text{time}(\zeta)=\zeta\xi(\zeta)$ is
\beq \label{EM::inv-corr-time}
    \frc1{\xi^{\rm time}}
    =
    -\int_{-\pi}^\pi
    \frac{dp}{2\pi}
    |\zeta^{-1}-\ep'(p)|\log\left|\tanh\frac{\beta\ep(p)}2\right|.
\eeq
We note that $\ep'(p) = h\sin( p)/\ep(p)$, and $\p_h \ep(p) = (h-\cos p)/\ep(p)$. The function $\ep(p)$ is strictly positive for all $(p,h)\neq (0,1)$ ($p\in[-\pi,\pi], h>0$), so the divergence can only come from the region near $p=0,h=1$. For $(p,h)\neq (0,1)$ we can exchange $p$-integral and $h$-derivative. Writing $h = 1+\delta$, we evaluate the $h$ derivative of the integrand $I(p,h)$  
and take its leading order in $(p,\delta)$ around $(0,0)$. We use $\ep(p) \sim \sqrt{p^2 + \delta^2}$, neglecting higher order in $\delta, p$ (with $\sqrt{p^2 + \delta^2}$ being of order 1 in both variables). In taking the $h$ derivative of the integrand, two terms appear, from $\p_h |\zeta^{-1}-\ep'(p)|$ and $\p_h \log|\tanh\beta\ep(p)/2|$. The first has the leading form
\beq\label{former}
\Big(\frc{d I}{dh}\Big)_1 \sim -\frc{\delta\,p\,\mathrm{sgn}(\zeta^{-1}-v(p))}{(p^2+\delta^2)^{3/2}}\log(\sqrt{p^2+\delta^2}\beta/2)\,,
\eeq
where $v(p)=p/\sqrt{p^2+\delta^2}$. The integral of the second term,
\beq\label{latter}
\Big(\frc{d I}{dh}\Big)_2 \sim
-\frc{\delta|\zeta^{-1}-v(p)|}{p^2+\delta^2}\,,
\eeq
is finite as $\delta\to0$, so we can focus on the first term. In order to extract the divergence of the $p$-integral of \eqref{former}, we may extend the integral to $\int_{-\infty}^\infty dp$, since the divergence as $\delta\to0$ comes from the region $p\sim 0$. The integration region should be split into two parts according to the sign function in Eq. \eqref{former}: $[-\infty,p_0]$ and $[p_0,\infty]$ with $p_0=\delta/\sqrt{\zeta^2-1}$.
We can also drop the $\beta/2$ factor in the log as it gives a subleading finite contribution. Then using
\begin{equation}
\int\dd p \frc{p}{(p^2+\delta^2)^{3/2}}\log(p^2+\delta^2)
= \frc{d}{da}\int_{p_1}^{p_2} \dd p\,
    \frc{p}{(p^2+\delta^2)^{3/2-a}}\Bigg|_{a=0}    =
   \frc{d}{da} \left.\left[\frc{(p^2+\delta^2)^{a-1/2}}{2a-1}\right]_{p_1}^{p_2} \right|_{a=0} 
\end{equation}
we obtain

\begin{equation}
\begin{split}
    \int_{-\pi}^\pi\frc{dp}{2\pi}
    \Big(\frc{d I}{dh}\Big)_1
    &\sim
    \frc{\delta}2
    \left(\int_{-\infty}^{p_0}\frc{dp}{2\pi}-\int_{p_0}^\infty\frc{dp}{2\pi}\right)
    \,
    \frc{p}{(p^2+\delta^2)^{3/2}}\log(p^2+\delta^2)\\
    &=
    \frc{\delta}{4\pi}
    \frc{d}{da}\left(2\frac{(p_0^2+\delta^2)^{a-1/2}}{2a-1}
   \,
    \right)\Bigg|_{a=0}
    =-\frc{\delta}{2\pi}\frac{2+\log(p_0^2+\delta^2)}{(p_0^2+\delta^2)^{1/2}}
    \sim
    -\frac{\sgn(\delta)}\pi\frc{\sqrt{\zeta^2-1}}{\zeta}\log |\delta|.
\end{split}
\end{equation}

Similarly, we can prove Eq.\ (\ref{eq::fluctuation-derivative-wrt-zeta}) by analyzing the behavior of (\ref{EM::inv-corr-time}) at $h = 1$ and $\zeta = 1 + \varphi$ for $\varphi \ll 1$:
\begin{equation}
    \frac{1}{\zeta} = \frac{\zeta}{\xi^\text{time}} = -\int_{-\pi}^\pi \frac{dp}{2\pi} \left | 1-\frac{(1+\varphi )\sin p}{\sqrt{2-2\cos p}} \right | \log \left | \tanh \frac{\beta \sqrt{2 - 2 \cos p}}{2}\right |.
\end{equation}
The expression inside the first absolute value is negative if and only if $0 < p < p_*$, where $p_* = \sqrt{8 \varphi}$. Then, we can forget about this absolute value if we subtract this region twice:
\begin{equation}
    \frac{1}{\xi(\zeta = 1+\varphi)} = \frac{1}{\xi(\zeta = 1)} + 2 \int_{0}^{\sqrt{8 \varphi}} \frac{dp}{2\pi} \left (1-\frac{(1+\varphi) \sin k}{\sqrt{2 - 2 \cos k}} \right )\log \left | \tanh \frac{\beta \sqrt{2 - 2 \cos p}}{2}\right |,
\end{equation}
where the first term appears by noticing that the $\varphi$-dependent part of the integrand without the absolute value is odd in $p$. To calculate the second term, we can expand the integrand to second order in $k$ and integrate the result exactly:
\begin{equation}
    \frac{1}{\xi(\zeta = 1+\varphi)} = \frac{1}{\xi(\zeta = 1)} - \frac{2 \sqrt{2}}{9 \pi}(-8 + 3 \log( 2\varphi \beta^2)) \varphi^{3/2}.
\end{equation}
Taking the derivative with respect to $\varphi$ yields Eq.\ (\ref{eq::fluctuation-derivative-wrt-zeta}).

\paragraph{Analytic expressions.---}

In the regime $1<h<1/\zeta$, the asymptotic behavior of the correlation function is governed by the interplay of poles and saddle points in the fermionic integral \eqref{eq:fermionic_part}. For the sake of completeness, we provide here explicit analytic expressions that determine the correlation length. 

The fermionic contribution to the inverse correlation length is 
\begin{equation}
    \xi_\text{pr}^{-1} = \max \{\Phi^{(1)}(z_{P0}),\,\dots,\Phi^{(1)}(z_{Pn_\text{max}}),\,\Phi^{(1)}(z_{S1})\}\,,
\end{equation}
where $\Phi^{(1)}(z) = i z - i \epsilon(z) \zeta$ and $n_\text{max}$ is the number of poles inside the contour:
\begin{equation}
    n_{\text{max}} = \max \left \{n\; |\; n \in \mathbb{N},\; \text{Im}\,z_{Pn}<\text{Im}\,z_{S1} \right \}.
\label{eq:EM:nmax}
\end{equation}
The poles are located at
\begin{equation}
        z_{Pn} = \arccos \left ( \frac{(1+h^2)\beta^2  + 4 \pi^2 n^2}{2h\beta^2}\right ) \in i \mathbb{R}_+, \quad n=0,1,\dots
    \label{eq:EM:pole-positions}
    \end{equation}
The saddle point relevant for the calculation is located at $z_{S1} \in i \mathbb{R}_+$ satisfying 
    \begin{equation}
        \cos z_{S1} = \frac{1+\sqrt{(1-\zeta^2)(1-\zeta^2 h^2)}}{h \zeta^2}\,.
    \label{eq:EM:saddle-point1}
    \end{equation}
We note that the exponent function $\Phi^{(1)}(z)$ evaluated at the poles is a real number that can be given explicitly:
    \begin{equation}
        \Phi^{(1)}(z_{Pn}) = \frac{2\pi n \zeta}{\beta} - \mathrm{arccosh} \left ( \frac{(1+h^2)\beta^2  + 4 \pi^2 n^2}{2h\beta^2}\right )\,. 
    \label{eq:EM:pole-exponent-values}
    \end{equation}


\begin{center}
    \large{\bf \underline{Supplementary Material}\\
``Observing quantum phase transitions at nonzero temperature:\\
non-analytic behavior of order-parameter correlation times''
}
\end{center}

In this Supplementay Material, we provide some technical details that were mentioned in the main body of the Letter. In particular, we will :
\begin{enumerate}
    \item summarize the diagonalization of the transverse-field Ising model;
    \item explain how the factorization of the correlation function in Eq. (8) of the main text is achieved;
    \item we evaluate the correlation length $\xi$ in the various regimes of the model.
\end{enumerate}

\section{Diagonalization of the Transverse-Field Ising Model}
This section provides an overview of the diagonalization and the Hilbert space structure of the transverse-field Ising model (TFIM) with periodic boundary conditions \cite{LSM,CEF}. We discuss the Jordan–Wigner transformation and the Bogoliubov rotation to map the model onto free fermions. We pay special attention to the differences between the parity-even Neveu–Schwarz (NS) and parity-odd Ramond (R) sectors and their implications for the spectrum in the ferromagnetic phase.

The Hamiltonian under consideration can be written as
    \begin{equation}
        H = -\frac{J}{2} \sum_{j = 1}^L \sigma_j^1 \sigma_{j+1}^1 + h \sigma_j^3,
        \label{eq:SM:Ising-Hamiltonian}
    \end{equation}
where $J>0$ gives the energy scale of the system, $L$ is the number of lattice sites, $h$ is the strength of the transverse magnetic field and $\sigma_j^\alpha$ are the Pauli matrices on site $j$. We impose periodic boundary conditions: $\sigma_{L+1}^\alpha = \sigma_1^\alpha$. Using the Jordan–Wigner transformation \cite{JW}, we can introduce the fermionic creation and annihilation operators $c_j$ and $c_j^\dagger$, and rewrite the Pauli matrices as
    \begin{equation}
        \sigma_j^3 = 1-2c_j^\dagger c_j,\qquad \qquad \sigma_j^1 = \prod_{l = 1}^{j-1} \left ( 1-2c_l^\dagger c_l \right )\left(c_j + c_j^\dagger\right).
    \label{eq:SM:JW-trafo}
    \end{equation}    
Note that the inverse transformations are 
    \begin{equation}
        c_j = \prod_{l = 1}^{j-1}\left(\sigma_l^3\right) \sigma_j^+, \qquad c_j^\dagger = \prod_{l = 1}^{j-1}\left (\sigma_l^3\right) \sigma_j^-,
        \label{eq:SM:JW-inverse}
    \end{equation}
 where $\sigma_j^\pm = \frac{1}{2}(\sigma_j^1 \pm i \sigma_j^2)$. In this language, the Hamiltonian reads
    \begin{equation}
        H = -\frac{J}{2}\left [\sum_{j = 1}^{L-1} \left ( (c_j^\dagger -c_j)(c_{j+1}+c_{j+1}^\dagger)\right )+e^{i \pi N}(c_L - c_L^\dagger)(c_1^\dagger + c_1)+ \sum_{j = 1}^L h (1-2 c_j^\dagger c_j)\right ],
    \label{eq:SM:H-JW-coordinate-space}
    \end{equation}
where 
    \begin{equation}
        e^{i \pi N}  = e^{i\pi\sum_{l = 1}^L c_l^\dagger c_l} = \prod_{l = 1}^L (1-2c_l^\dagger c_l)
    \label{eq:SM:parity-with-c} 
    \end{equation}
is the parity operator, which is a conserved quantity since it commutes with the Hamiltonian: $[H,e^{i \pi N}] = 0$.   Therefore, we can choose an eigenbasis of $H$ that is block-diagonal with respect to the even and odd subspaces:
    \begin{equation}
        H = P_e H_e P_e + P_o H_o P_o, \qquad  P_{e/o} =\frac{1}{2} \left ( 1\pm e^{i \pi N} \right ).
    \label{eq:SM:block-diagonal-H}
    \end{equation}
Using Eqs.\ (\ref{eq:SM:H-JW-coordinate-space}) - (\ref{eq:SM:block-diagonal-H}), the even and odd Hamiltonians take the same form
    \begin{equation}
        H_{e/o} = -\frac{J}{2}\sum_{j = 1}^L \left ( (c_j^\dagger -c_j)(c_{j+1}+c_{j+1}^\dagger) + h (1-2 c_j^\dagger c_j)\right )
    \label{eq:SM:even-odd-H-coordiante-space}
    \end{equation}
with the difference manifesting in the boundary conditions of the fermions: 
    \begin{equation}
    \begin{aligned} 
        c_{L+1} &= -c_1\qquad \text{in}\; \; H_e, \qquad (\text{NS sector})\\
        c_{L+1} &= c_1\;\;\; \qquad\text{in}\;\; H_o. \qquad (\text{R sector})
    \label{eq:SM:bondary-conditions-on-c}
    \end{aligned}
    \end{equation}

Using the translational invariance of Eqs.\ (\ref{eq:SM:even-odd-H-coordiante-space}) and (\ref{eq:SM:bondary-conditions-on-c}) we  convert the fermions to Fourier space:
    \begin{equation}
        c_j = \frac{1}{\sqrt{L}}\sum_{n \in S_L}c_{p_n} e^{-i p_n j}= \frac{1}{\sqrt{L}}\sum_{n\in S_L}c_{q_n} e^{-i q_n j},
    \label{eq:SM:real-to-Fourier}
    \end{equation}
where the two boundary conditions in the two sectors impose the following momentum quantization:
    \begin{equation}
    \begin{aligned}
        p_n= \frac{2\pi n}{L} + \frac{\pi}{L},\qquad q_n = \frac{2\pi n}{L}, \qquad S_L = \left \{-\lfloor L/2\rfloor,\; -\lfloor L/2\rfloor+1,\;\dots \lceil L/2\rceil-1 \right \}.
    \label{eq:SM:momentum-quantization}
    \end{aligned}
    \end{equation}
The inverse transformations are simply given by
    \begin{equation}
        c_{p_n} = \frac{1}{\sqrt{L}}\sum_{j = 1}^L c_j e^{i p_n j},\qquad c_{q_n} = \frac{1}{\sqrt{L}}\sum_{j = 1}^L c_j e^{i q_n j}.
    \label{eq:SM:Fourier-to-real}
    \end{equation}
Rewriting $H_\text{e(o)}$ and $e^{i \pi N}$ in terms of the $c_{p_n}$( $c_{q_n})$ variables, we get 
    \begin{equation}
    \begin{aligned}
        H_e &= -\frac{J}{2} \left [\sum_{n \in S_L} 2(\cos p_n-h)c_{p_n}^\dagger c_{p_n} + c_{p_n}^\dagger c_{-p_n}^\dagger e^{-ip_n} - c_{p_n}c_{-p_n}e^{i p_n}+h-e^{ip_n}\right ],\\
        H_o &= -\frac{J}{2} \left [\sum_{n \in S_L} 2(\cos q_n -h)c_{q_n}^\dagger c_{q_n} + c_{q_n}^\dagger c_{-q_n}^\dagger e^{-iq_n} - c_{q_n}c_{-q_n}e^{i q_n}+h-e^{iq_n}\right ],\\
        &\qquad \qquad \quad e^{i \pi N} = \prod_{n \in S_L} (1-2 c_{p_n}^\dagger c_{p_n})= \prod_{n \in S_L} (1-2 c_{q_n}^\dagger c_{q_n}).
    \label{eq:SM:even-odd-H-momentum-space}
    \end{aligned}
    \end{equation}
Note that $-p_n$ and $-q_n$ always exist, because momenta are understood modulo $2\pi$. 

The appearance of terms violating fermion-number conservation indicates that a Bogoliubov transformation is necessary to complete the diagonalization. Let us first consider the even sector:
    \begin{equation}
    \begin{aligned}
        c_{p_n} = \cos(\theta_{p_n}/2) \alpha_{p_n} + i \sin (\theta_{p_n}/2) \alpha_{-p_n}^\dagger,\\
        \alpha_{p_n} = \cos(\theta_{p_n}/2) c_{p_n} - i \sin (\theta_{p_n}/2) c_{-p_n}^\dagger.
    \label{eq:SM:Bogoliubov-transformations}
    \end{aligned}
    \end{equation}
Inserting this into $H_{e}$, the terms with two creation or annihilation operators cancel if and only if the Bogoliubov angle satisfies
    \begin{equation}
        \tan\theta_{p_n} = \frac{\sin p_n}{\cos p_n -h}.
    \label{eq:SM:Bololiubov-angle-tan}
    \end{equation}
This has two solutions for the cosine of the Bogoliubov angle:
    \begin{equation}
        \cos \theta_{p_n} = \delta_{p_n} \frac{h-\cos p_n}{\sqrt{1+h^2-2h\cos p_n}},
        \label{eq:SM:Bogoliubov-angle-cos}
    \end{equation}
where $\delta_{p_n}$ can be $+1$ or $-1$. Substituting Eqs.\ (\ref{eq:SM:Bogoliubov-transformations}) and (\ref{eq:SM:Bogoliubov-angle-cos}) into $H_e$, we get
    \begin{equation}
        H_e =  \sum_{n \in S_L} J \delta_{p_n}\sqrt{1+h^2-2h\cos p_n}\left (\alpha_{p_n}^\dagger \alpha_{p_n}-\frac{1}{2}\right).
    \label{eq:SM:He-in-Bogolibov-space}
    \end{equation}
This is the diagonal form of the Hamiltonian in the even sector. We can see that the choice of $\delta_{p_n}$ represents a particle-hole transformation between $\alpha_{p_n}$ and $\alpha_{p_n}^\dagger$. Furthermore, even if we require (as we do in this work) that the even subspace projector $P_e$ projects to states with an even number of $\alpha_{p_n }^\dagger$ excitations,
    \begin{equation}
        e^{i\pi N} = \prod_{n \in S_L} (1-2 c_{p_n}^\dagger c_{p_n}) = \prod_{n \in S_L} (1-2\alpha_{p_n}^\dagger \alpha_{p_n}),
    \label{eq:SM:parity-Bogoliubov-language}
    \end{equation}
it does not give any additional constraints on $\delta_{p_n}$, since
    \begin{equation}
        (1-2c_{p_n}^\dagger c_{p_n})(1-2c_{-p_n}^\dagger c_{-p_n}) = (1-2\alpha_{p_n}^\dagger \alpha_{p_n})(1-2\alpha_{-p_n}^\dagger \alpha_{-p_n}),
    \label{eq:SM:even-sector-parity-checl}
    \end{equation}
holds for any choice of $\delta_{p_n}$ (even for those edge cases possible for odd $L$, where for $n = \lceil{L/2}\rceil-1$, $p_n = \pi = -p_{n} \;\; \text{mod} \;2\pi$).

The calculation is almost identical in the odd sector: Eqs.\ (\ref{eq:SM:Bololiubov-angle-tan})-(\ref{eq:SM:even-sector-parity-checl}) are still valid if we change $p_n$ to $q_n$ and $H_e$ to $H_o$. However, the transformation of the parity operator requires a special treatment of the zero momentum mode $q_0 = 0$, since $q_0 = -q_0$. Indeed, if we require
    \begin{equation}
        e^{i\pi N}= \prod_{l = 1}^L (1-2 c_{l}^\dagger c_{l}) = \prod_{n \in S_L} (1-2\alpha_{q_n}^\dagger \alpha_{q_n}),
    \label{eq:SM:odd-parity-rquirement-Bogoliubov-language}
    \end{equation}
in addition to the analog of Eq.\ (\ref{eq:SM:even-sector-parity-checl}), we also need to satisfy
    \begin{equation}
        (1-2c_0^\dagger c_0) = (1-2 \alpha_0^\dagger \alpha_0).
        \label{eq:SM:zero-mode-parity-requiremet}
    \end{equation}
This dictates $\cos \theta_{q_0} = 1$, which fixes $\delta_{q_0} = \text{sign}(h-1)$ according to Eq. (\ref{eq:SM:Bogoliubov-angle-cos}). In the end, we get
    \begin{equation}
        H_o =  \sum_{n \in S_L, n \neq 0} J\delta_{q_n}\sqrt{1+h^2-2h\cos q_n}\left (\alpha_{q_n}^\dagger \alpha_{q_n}-\frac{1}{2}\right) +J (h-1) \left (\alpha_0^\dagger \alpha_0-\frac{1}{2} \right ),
    \label{eq:SM:Ho-in-Bogolibov-space}
    \end{equation}
where the remaining $\delta_{p_n}$ signs can be chosen freely (keeping in mind that the different choices are equivalent to a particle-hole transformation).

Since we are free to choose the remaining $\delta$ signs appearing in Eqs.\ (\ref{eq:SM:He-in-Bogolibov-space}) and (\ref{eq:SM:Ho-in-Bogolibov-space}), we use the regular convention where all of them are $+1$, so $\alpha_{p_n}^\dagger$ and $\alpha_{q_n\neq0}^\dagger$ creates an excitation with positive energy:
    \begin{equation}
    \begin{aligned}
        H_e =  \sum_{n \in S_L} \epsilon_{p_n}\left (\alpha_{p_n}^\dagger \alpha_{p_n}-\frac{1}{2}\right), &\qquad 
        H_o =  \sum_{n \in S_L
        } \epsilon_{q_n}\left (\alpha_{q_n}^\dagger \alpha_{q_n}-\frac{1}{2}\right),\\
        e^{i\pi N} = \prod_{n \in S_L} (1-2\alpha_{p_n}^\dagger& \alpha_{p_n}) = \prod_{n \in S_L} (1-2\alpha_{q_n}^\dagger \alpha_{q_n}),
    \label{eq:SM:diagonalization-end-result}
    \end{aligned}
    \end{equation}
where  
    \begin{equation}
        \epsilon_{p_n} = \epsilon(p_n),\qquad \qquad 
        \epsilon_{q_n} = \begin{cases}\epsilon(q_n)\;\quad\qquad\text{if}\;\; n \neq 0\\
        J(h-1),\;\quad \text{if}\;\;n = 0,\\
        \end{cases}
    \label{eq:SM:free-fermion-energies}
    \end{equation}
with
    \begin{equation}
        \epsilon(p) = J\sqrt{1+h^2 - 2h \cos p}
    \label{eq:SM:dispersion-relation}
    \end{equation}
giving the dispersion relation in the continuum. Note that $\epsilon_{q_0} = \text{sgn}(h-1) \,\epsilon(0).$

Before we conclude with the general discussion of the model, we have to discuss the structure of the Hilbert space. First, we can define the vacuum state $\ket{\mathbf{0}}_{\text{NS}}$ that is annihilated by all $\alpha_{p_n}$ operators that diagonalize the even sector of the Hilbert space. It has the following properties:
    \begin{equation}
        e^{i \pi N} \ket{\mathbf{0}}_{\text{NS}} = \ket{\mathbf{0}}_{\text{NS}}, \qquad H\ket{\mathbf{0}}_{\text{NS}} = H_e \ket{\mathbf{0}}_{\text{NS}} = E_\text{NS}\ket{\mathbf{0}}_{\text{NS}}, \qquad E_\text{NS} =-\frac{1}{2} \sum_{n \in S_L} \epsilon_{p_n }.
    \label{eq:SM:true-ground-state-properties}
    \end{equation}
Note that $H_o \ket{0}$ is something highly non-trivial, but by Eq.\ (\ref{eq:SM:block-diagonal-H}) this term does not contribute to the energy. We can build the Fock states by acting with an even number of creation operators:
    \begin{equation}
        \ket{p_1, p_2, \dots p_{2j}} = \alpha_{p_1}^\dagger \alpha_{p_2}^\dagger \dots \alpha_{p_{2j}}^\dagger \ket{\mathbf{0}}_\text{NS}, \qquad H \ket{p_1, p_2, \dots p_{2j}} = \left (\sum_{i = 1}^{2j} \epsilon_{p_{i}} +E_\text{NS}\right )\ket{\mathbf{0}}_\text{NS}.
    \label{eq:SM:even-Fock-states}
    \end{equation}
These states are orthonormal and form a complete basis on the even sector of the Hilbert space. 

We can also define $\ket{\mathbf{0}}_\text{R}$ as the state annihilated by all $\alpha_{q_n}$ operators:
    \begin{equation}
        e^{i \pi N} \ket{\mathbf{0}}_\text{R} = \ket{\mathbf{0}}_\text{R}, \qquad  H_o \ket{\mathbf{0}}_\text{R} = E_\text{R} \ket{\mathbf{0}}_\text{R}, \qquad E_\text{R} = -\frac{1}{2}\sum_{n \in S_L} \epsilon_{q_n},
    \label{eq:SM:Ramond-vacuum-properties}
    \end{equation}
but it is not an energy eigenstate, since it has positive parity (cf.\ Eq.\ (\ref{eq:SM:block-diagonal-H})). In fact, we can express it in principle as a (large) linear combination of basis states from the even sector. However, if we act on it with an odd number of $\alpha_{q_n}^\dagger$ creation operators, we get the energy eigenstates of the odd sector:
    \begin{equation}
        \ket{q_1,q_2,\dots q_{2j+1}} = \alpha_{q_1}^\dagger \alpha_{q_2}^\dagger \dots \alpha_{q_{2j+1}}^\dagger \ket{\mathbf{0}}_\text{R}, \qquad H \ket{q_1,q_2,\dots q_{2j+1}} = \left ( \sum_{i = 1}^{2j+1} \epsilon_{q_{i}} +E_\text{R} \right ) \ket{\mathbf{0}}_\text{R}.
    \label{eq:SM:odd-Fock-states}
    \end{equation}
These states form an orthonormal basis on the odd subspace of the Hilbert space.

It is also useful to define the state with a zero momentum excitation $\ket{\Delta} = \alpha_{q_0}^\dagger \ket{\mathbf{0}}_\text{R}$, since in the ferromagnetic phase, its energy becomes exponentially close to the ground state energy $E_\text{NS}$:
    \begin{equation}
        H \ket{\Delta} = E_\Delta\ket{\Delta}\,, \qquad E_\Delta = (J (h-1) + E_\text{R})\, , \qquad E_\Delta - E_{\text{NS}} \asymp e^{-L \delta} \;\; \text{for}\;\; h<1\,.
    \label{eq:SM:ferromagnetic-degenerate-state}
    \end{equation}

\section{The doubling trick}
\label{app:doubling}

In this section, we discuss the details of the doubling trick that relates two copies of the quantum Ising chain to a discretized Dirac Hamiltonian. We prove that this procedure embeds the $\mathbb{Z}_2$ symmetry of the Ising chain into a continuous $U(1)$ symmetry. We start from the Hamiltonians $H_{e/o}$ in the even / odd sectors, Eq.~\eqref{eq:SM:even-odd-H-coordiante-space}, choosing a sector and dropping the subscript for lightness of notation:
    \begin{equation}
        H = 
        -\frac{J}{2}\sum_{j = 1}^L \left[(c_j^\dagger-c_j)(c_{j+1} + c_{j+1}^\dagger) +h(c_j - c_j^\dagger)(c_j + c_j^\dagger)\right]\,.
    \end{equation}
Recall that the sector only specifies the boundary conditions on $c_j$'s, and this does not affect the argument below. To simplify the formulas, we introduce the Majorana fermions (which satisfy the same boundary conditions as the original fermions)
    \begin{equation}
        a_j = c_j + c_j^\dagger,\qquad \bar{a}_j = i (c_j - c_j^\dagger)
    \end{equation}
that satisfy
    \begin{equation}
        a_j^2 = \bar{a}_j^2 = 1,\qquad a_j^\dagger = a_j,\qquad \bar{a}_j^\dagger = \bar{a}_j,\qquad \{a_j,a_k\} = \{\bar{a}_j,\bar{a}_k\} =2 \delta_{j,k},\qquad \{a_j,\bar{a}_k\} = 0\,.
    \end{equation}
In this language, the Hamiltonian becomes
    \begin{equation}
        H = -\frac{iJ}{2}\sum_{j = 1}^L \left(\bar{a}_j a_{j+1}-h \bar{a}_j a_j\right)\,.
    \end{equation}

Now consider two independent, anti-commuting copies with fermion operators $c_l,c_l^\dagger$ and $d_l,d_l^\dagger$ (satisfying the same boundary conditions) and corresponding Majorana operators $a_l,a_l^\dagger$, $b_l,b_l^\dagger$, and consider the direct product Hilbert space built on the tensor product of vacua intruduced in Eqs.\ (\ref{eq:SM:true-ground-state-properties}) and (\ref{eq:SM:Ramond-vacuum-properties}). The doubled model can be written as 
    \begin{equation}
        H^\text{d}=-\frac{iJ}{2}\sum_{j = 1}^L\left( \bar{a}_j a_{j+1}-h \bar{a}_j a_j +\bar{b}_j b_{j+1}-h \bar{b}_j b_j\right)\,,
    \label{eq:doubledH}
    \end{equation}
where the two copies anti-commute:
    \begin{equation}
        \{a_j,b_k\} = \{\bar{a}_j,b_k\}=\{a_j,\bar{b}_k\}=\{\bar{a}_j,\bar{b}_k\}=0\,.
    \end{equation}
Next, we can introduce discrete derivatives (that are defined to handle the boundary conditions of $a_j$ and $b_j$) as
    \begin{equation}
        a_{j+1} = a_j + \partial_j a_j\,,\qquad b_{j+1} = b_j + \partial_j b_j
    \end{equation}
and rewrite the Hamiltonian into the form
    \begin{equation}
        H^\text{d} = -\frac{iJ}{2}\sum_{j = 1}^L \bar{a}_j \partial_j a_j+\bar{b}_j \partial_j b_j + (1-h) \bar{a}_j a_j + (1-h)\bar{b}_j b_j\,.
    \end{equation}
Now we can introduce the Dirac fermions 
    \begin{equation}
        \Psi_j =\frac{1}{2} \begin{pmatrix} a_j + i b_j \\ \bar{a}_j + i \bar{b}_j
        \end{pmatrix}, \qquad \Psi_j^\dagger =\frac{1}{2} \begin{pmatrix} a_j - i b_j & \bar{a}_j - i \bar{b}_j
        \end{pmatrix}
    \end{equation}
as well as the $\gamma$-matrices
    \begin{equation}
        \gamma^0 = \begin{pmatrix}0 & i \\ -i & 0\end{pmatrix},\qquad \gamma^1 = \begin{pmatrix}i & 0\\ 0 &-i\end{pmatrix}
    \end{equation}
that satisfy the Clifford-algebra
    \begin{equation}
        \left(\gamma^0\right)^2 = 1,\qquad \left(\gamma^1\right)^2 = -1,\qquad \{\gamma^0,\gamma^1\} = 0\,.
    \end{equation}
Then, the Hamiltonian can be written in the Dirac form
    \begin{equation}
        H^\text{d} = J\sum_j \Psi_j^\dagger \left ( -i \gamma^0 \gamma^1 \partial_j + m \gamma^0 \right) \Psi_j\,.
    \end{equation}
    
Unlike the original Ising spin chain, this model has a conserved $U(1)$ charge 
    \begin{equation}
        \tilde Q = \sum_{l = 1}^L \Psi_l^\dagger \Psi_l = \sum_{l = 1}^L \Big(1 + \frac{i}{2}a_l b_l + \frac{i}{2}\bar{a}_l \bar{b}_l\Big)\,.
    \label{eq:U(1)charge}
    \end{equation}
It can be checked by explicit calculation that it commutes with $H$. It is more convenient to remove the contribution of the identity from the charge and rephrase it as
    \begin{equation}
        Q = \tilde Q - L = \sum_{l = 1}^L q_l = \sum_{l = 1}^L \Big(i c_l^\dagger d_l -i d_l^\dagger c_l\Big)\,,
    \end{equation}
where $q_l$ is the charge density at site $l$. We can study it in the local basis of the $c_l$ and $d_l$ fermions, $ 1 \rightarrow\ket{0}  \ket{0},\; 2 \rightarrow \ket{1} \ket{0},\;3 \rightarrow\ket{0} \ket{1},\;4 \rightarrow\ket{1} \ket{1},$
where its matrix form becomes    \begin{equation}
        q_l = \begin{pmatrix}
            0&0&0&0\\0&0&-i&0\\0&i&0&0\\0&0&0&0
        \end{pmatrix}\,.
    \end{equation}
Therefore, it is easy to calculate the vertex operator
    \begin{equation}
        e^{i \alpha q_l} = \begin{pmatrix}
            1&0&0&0\\0 & \cos \alpha & \sin \alpha &0\\0 &-\sin \alpha & \cos \alpha &0\\0&0&0&1
        \end{pmatrix},
    \end{equation}
which becomes diagonal for $\alpha = \pi$:
    \begin{equation}                                                  e^{i \pi q_l} = \begin{pmatrix}
            1&0&0&0\\0&-1&0&0\\0&0&-1&0\\0&0&0&1
        \end{pmatrix}.
    \end{equation}
This allows us to show that it factorizes between the copies in the following way:
     \begin{equation}
        e^{i \pi q_l}= (1-2c_l^\dagger c_l)(1-2 d_l^\dagger d_l) = (-i\bar a_la_l)(-i\bar b_lb_l).
        \label{eq:localvertexnormalordered}
    \end{equation}
and in particular we note that the resulting operator is normal-ordered with respect to $c_l,d_l,c^\dagger_l,d^\dagger_l$'s.
We can also construct the twist field operator
    \begin{equation}
         \mathcal U_j^\pi =\,
         :e^{i\pi \sum_{l=1}^{j-1} \Psi_l^\dagger \Psi_l}:\,=e^{i\pi \sum_{l=1}^{j-1} q_l}=\prod_{l=1}^{j-1}(1-2c_l^\dagger c_l)\prod_{l=1}^{j-1}(1-2d_l^\dagger d_l)= (-1)^{(j-1)}\prod_{l=1}^{j-1}(\bar a_la_l)(\bar b_lb_l),
         \label{eq:SM:twistU}
    \end{equation}
which, crucially, by the third equality, is equal to the product of the Jordan--Wigner strings in the two copies. Note that it satisfies the $\mathbb Z_2$-twist-field exchange relations \cite{e27121230}
    \begin{equation}
        \mathcal U_j^\pi (c_l , c_l^\dagger, d_l , d_l^\dagger)
        = (c_l,c_l^\dagger,d_l , d_l^\dagger)\mathcal U_j^\pi \ (l\geq j)\,,\qquad-(c_l,c_l^\dagger,d_l , d_l^\dagger)\mathcal U_j^\pi \ (l< j)\,.
    \end{equation}

\section{Factorization of the correlation function}
\label{app:factor}

In this section, we show the factorization properties of the finite temperature correlation functions given in Eq.\ (8) in the main text. First, let us recall the definition of these correlators:
    \begin{equation}
        \braket{\sigma_1^1(0) \sigma_{j}^1(t)} = \frac{\mathrm{Tr} \left ( e^{-\beta H} \sigma_1^1(0) \sigma_j^1(t)  \right )}{\mathrm{Tr} \left ( e^{-\beta H}\right )}, 
    \label{eq:SM:finite-T-corr-func-def}
    \end{equation}
were the time evolution of the order parameter is understood in the Heisenberg picture:
    \begin{equation}
    \sigma^1(t) = e^{i H t} \sigma^1 e^{-i H t}.
    \label{eq:SM:Hesienberg-evolution}
    \end{equation}
Note that we placed the first operator on the first site at zero time, which can be done without loss of generality.

Let us first focus on the even sector; the odd sector works similarly.   
According to the Jordan--Wigner transformation (\ref{eq:SM:JW-trafo}), 
\begin{equation}
    \sigma_j^1 = \prod_{l=1}^{j-1} \left ( 1-2 c_l^\dagger c_l \right )\left ( c_j + c_j^\dagger\right ) = \prod_{l=1}^{j} \left ( 1-2 c_l^\dagger c_l \right )\left ( c_j - c_j^\dagger\right ) = 
    \prod_{l=1}^{j}(-i\bar a_la_l)\,(-i\bar a_j)\,,
\end{equation}
where we extended the string by one site at its right end.
Then the product of operators can be written in terms of the Majorana fermions  
as
    \begin{equation}
        \sigma_1^1(0) \sigma_j^1(t) = 
          a_1 e^{iH_o t} \prod_{l=1}^{j}(-i\bar a_la_l)\,(-i\bar a_j)e^{-iH_e t}
          = a_1 e^{iH_o t}e^{-iH_e t}(-i)^{j+1}\left(\prod_{l=1}^{j}(\bar a_l(t)a_l(t)\right)\,\bar a_j(t)\,.
    \label{eq:SM:equal-time-operator-expansion}
    \end{equation}
where the fermionic time evolution is defined in a sector-dependent way, $(a_l(t),\bar a_l(t)) = e^{iH_{\rm e}t}(a_l,\bar a_l)e^{-iH_{\rm e}t}$, in analogy with Ref. \cite{XXmodel}.

In order to proceed using the BFT techniques of Refs. \cite{Myers1,Myers2}, we need a conserved $U(1)$ charge. To this end, we transition to the Dirac theory by taking the product of the correlators in the two copies and interpret it as a four-point function in the doubled theory:
    \begin{equation}
(\braket{\sigma_1^1 \sigma_j^1(t)})^2 = \left\langle(\sigma_1^1)_1(\sigma_j^1(t))_1(\sigma_1^1)_2(\sigma_j^1(t))_2\right\rangle_\text{d} \,,
    \end{equation}
where the 1,2 subscripts label the copies and the``d'' subscript refers to the doubled theory.

Plugging in Eq. \eqref{eq:SM:equal-time-operator-expansion} and using Eq. \eqref{eq:SM:twistU}, this can be written as
    \begin{multline}
        \left\langle a_1e^{iH^{(1)}_o t}e^{-iH^{(1)}_e t} 
        (-i)^{j+1}\left(\prod_{l=1}^{j}(\bar a_l(t)a_l(t)\right)\,\bar a_j(t)
        \,b_1 e^{iH^{(2)}_o t}e^{-iH^{(2)}_e t}
        (-i)^{j+1}\left(\prod_{l=1}^{j}(\bar b_l(t)b_l(t)\right)\,\bar b_j(t)
        \right\rangle\\
       =\left\langle a_1b_1 e^{iH^\text{d}_o t}e^{-iH^\text{d}_e t}\,\mathcal{U}_{j+1}^\pi(t) \,\bar a_j(t) \bar b_j(t)\right\rangle\,,
    \end{multline}
where we exploited that $H^{(1)}_{e/o}+H^{(2)}_{e/o}=H^\text{d}_{e/o}$.

Now we are in the position of invoking the results of Ref. \cite{XXmodel}. Generalizing the argument to the Dirac case, we obtain 
    \begin{equation}
        \braket{\sigma_1^1 \sigma_j^1(t)}^2 = -\braket{a_1 \bar a_j(t)b_1 \bar b_j(t)} _{i\pi}\braket{\mathcal{U}^\pi_{j+1}},
    \label{eq:SM:doubled-model-factorization}
    \end{equation}
where
    \begin{equation}
        \braket{a_1 \bar a_j(t)b_1 \bar b_j(t)} _{i\pi} = \frac{1}{\mathcal{Z}_\text{d}} \mathrm{Tr} \left (e^{-\beta H^\text{d} - i\pi Q} a_1 \bar a_j(t)b_1 \bar b_j(t)\right ).
    \end{equation}
is a grand canonical expectation value in the doubled theory at chemical potential associated to the $U(1)$ charge equal to $i\pi$.
According to Eqs.\ (\ref{eq:SM:parity-with-c}) and (\ref{eq:localvertexnormalordered}),
    \begin{equation}
        e^{-i \pi Q} = \prod_{l = 1}^L (1-2 c_l^\dagger c_l)(1-2 d_l^\dagger d_l) =   e^{-i \pi \sum_{l = 1}^L c_l^\dagger c_l}e^{-i \pi \sum_{l = 1}^L d_l^\dagger d_l} = e^{-i \pi N^{(1)}} e^{-i \pi N^{(2)}}\,.
    \end{equation}
This factorizes in the same way as the Hamiltonian does, so
    \begin{equation}
        \braket{a_1 \bar a_j(t)b_1 \bar b_j(t)} _{i\pi} =  \frac{\mathrm{Tr} \left (e^{-\beta H^{(1)}}e^{-i \pi N^{(1)}} a_1\bar a_j\right)}{\mathrm{Tr} \left ( e^{-\beta H^{(1)}} e^{-i\pi N^{(1)}}\right )}\frac{\mathrm{Tr} \left (e^{-\beta H^{(2)}}e^{-i \pi N^{(2)}} b_1\bar b_j\right)}{\mathrm{Tr} \left ( e^{-\beta H^{(2)}}e^{-i \pi N^{(2)}}\right )} = \braket{a_1\bar a_j(t)}_{i\pi} \braket{b_1\bar b_j(t)}_{i\pi}\,.
    \label{eq:SM:fermionic-part-factorization}
    \end{equation}
Note that in a single copy, the prescription $\mu=i\pi$ is translated into averages weighted by the fermionic parity operator.

Combining Eqs.\ (\ref{eq:SM:doubled-model-factorization}) and (\ref{eq:SM:fermionic-part-factorization}), we can write the factorization hypothesis in the single-copy TFIM model as
    \begin{equation}
        \braket{\sigma^1_1 \sigma_{1+x}^1(t)} \asymp \braket{a_1 \bar{a}_{1+x}(t)}_{i\pi} \braket{\mathcal{U}_{x+2}^\pi} \equiv F(x,t) \Omega(x,t),
    \label{eq:SM:factorization-hypothesis}
    \end{equation}
where $F(x,t)$ and $\Omega(x,t)$ is introduced to label the propagation and fluctuation contributions. In the remainder of this section, we will study the right-hand side of this equation in the $1 \ll x, Jt \ll L$, $\zeta = Jt/x$ fixed hydrodynamic limit. For simplicity, we will use the $J = 1$ convention.

Let us start with the fluctuation part. In the disordered, $h>1$ phase, we can use the same argument as in Ref.\ \cite{XXmodel} to conclude that 
    \begin{equation}
        \Omega_\text{d}(x,t) = \exp \left ( \int_{-\pi}^\pi \frac{dq}{2\pi} \left | x - \epsilon'(q) t \right | \log \tanh \left ( \frac{\beta \epsilon(q)}{2} \right )\right ),\;\; \text{for}\;\; h > 1.
        \label{eq:SM:string-part-for-paramagnet}
    \end{equation}
However, in the ordered phase, the zero mode $\ket{\Delta}$ causes a discontinuity in the spectrum, which gives rise to an extra contribution 
    \begin{equation}
        \Omega_\text{o}(x,t) =e^{-x \delta}\exp \left ( \int_{-\pi}^\pi \frac{dq}{2\pi} \left | x - \epsilon'(q) t \right | \log \tanh \left ( \frac{\beta \epsilon(q)}{2} \right )\right ), \;\; \text{for}\;\; h < 1.
        \label{eq:SM:string-part-for-ferromagnet}
    \end{equation}
Recall the definition of $\delta$ from Eq.\ (\ref{eq:SM:ferromagnetic-degenerate-state}). Note that such a time-independent prefactor also appears in the form-factor expansion of the continuum theory, cf.\ Ref.\ \cite{doyon2007finitetemperature}.

In order to study the fermionic contribution, we return to the original fermionic language,
    \begin{equation}
        F(x,t) = \frac{\mathrm{Tr} \left ( e^{-\beta H} e^{i \pi N} \left ( c_1 + c_1^\dagger \right )\left ( c_{1+x}(t) - c_{1+x}^\dagger(t) \right ) \right )}{\mathrm{Tr} \left ( e^{-\beta H} e^{i \pi N}\right )}\,,
    \label{eq:SM:fatorization-hypothesis-fermionic-part}
    \end{equation}
and evaluate it in the hydrodynamic limit. To calculate the appearing parity-modified thermal traces, we need to convert expression to the free fermionic modes:
    \begin{equation}
    \begin{aligned}
        \left ( c_1 + c_1^\dagger \right )\left ( c_{1+x}(t) - c_{1+x}^\dagger(t) \right ) &= \frac{1}{L} \sum_{n,m \in S_L} e^{i (p_n - p_m (1+x))} e^{-i\left ( \theta_{p_n} + \theta_{p_m}\right )/2} \left (\alpha_{p_n}^\dagger + \alpha_{-p_n}\right )\left (\alpha_{-p_m}^\dagger(t) + \alpha_{p_m}(t)\right ) = \\&= \frac{1}{L} \sum_{n,m \in S_L} e^{i (q_n - q_m (1+x))} e^{-i\left ( \theta_{q_n} + \theta_{q_m}\right )/2} \left (\alpha_{q_n}^\dagger + \alpha_{-q_n}\right )\left (\alpha_{-q_m}^\dagger(t) + \alpha_{q_m}(t)\right ).
    \end{aligned}
    \label{eq:SM:fermionic-part-derivation-free-fermion-formula}
    \end{equation}
Since the time-dependence of the fermions is understood in a sector-dependent way (cf.\ Eq.\ (\ref{eq:SM:equal-time-operator-expansion})),
    \begin{equation}
        \alpha_{p_m}(t) = e^{-i \epsilon_{p_m}t} \alpha_{p_m},\qquad \alpha_{q_m}(t) = e^{-i \epsilon_{q_m}t} \alpha_{q_m}.
    \label{eq:SM:time-evolution-approx}
    \end{equation}
Next, using Eq.\ (\ref{eq:SM:block-diagonal-H}), it is easy to check that 
    \begin{equation}
        e^{-\beta H} = P_\text{e} e^{-\beta H_\text{e}} P_\text{e} + P_\text{o} e^{-\beta H_\text{o}} P_\text{o}, \qquad [P_\text{e/o}, H_\text{e/o}] = [P_\text{e/o}, H_\text{o/e}] = 0,
    \label{eq:SM:exponential-H-block-diagonal-form}
    \end{equation}
which means that for all $\mathcal{O}$ operators,
    \begin{equation}
        \mathrm{Tr} \left ( e^{-\beta H} e^{i \pi N} \mathcal{O} \right ) = \mathrm{Tr} \left ( e^{-\beta H} e^{i \pi N} \mathcal{O} \right ) = \mathrm{Tr} \left (P_\text{e} e^{-\beta H_\text{e}} \mathcal{O} \right ) - \mathrm{Tr} \left (P_\text{o} e^{-\beta H_\text{o}} \mathcal{O} \right ).
    \label{eq:SM:modified-thermal-exp-value-of-general-operator}
    \end{equation}
Luckily, to calculate Eq.\ (\ref{eq:SM:fatorization-hypothesis-fermionic-part}), we can use the even (odd) sector representation in Eq.\ (\ref{eq:SM:fermionic-part-derivation-free-fermion-formula}) when evaluating the trace in the even(odd) sector, so we only need to consider terms, like
    \begin{equation}
        \mathrm{Tr} \left (P_\text{e} e^{-\beta H_\text{e}} \alpha^\dagger_{p_n} \alpha_{p_m} \right )\;\text{and}\;\mathrm{Tr} \left ( P_\text{o} e^{-\beta H_\text{o}} \alpha^\dagger_{q_n} \alpha_{q_m} \right ).
    \label{eq:SM:free-fermioic-exp-values-needed}
    \end{equation}
These are relatively easy to obtain because $H_{\text{e}(\text{o})}$ is built solely from $\alpha_{p_n} (\alpha_{q_n})$:
    \begin{equation}
    \begin{aligned}
        \mathrm{Tr} \left (P_\text{e} e^{-\beta H_\text{e}} \alpha^\dagger_{p_n} \alpha_{p_m} \right ) &= \frac{\delta_{n,m}}{2} \left ( \prod_{r \in S_L} 2 \cosh \left ( \frac{\beta \epsilon_{p_r}}{2}\right ) \frac{1}{1+e^{\beta \epsilon_{p_n}}} +  \prod_{r \in S_L} 2 \sinh \left ( \frac{\beta \epsilon_{p_r}}{2}\right ) \frac{1}{1-e^{\beta \epsilon_{p_n}}}\right ),\\ \mathrm{Tr} \left (P_\text{o} e^{-\beta H_\text{o}} \alpha^\dagger_{q_n} \alpha_{q_m} \right ) &= \frac{\delta_{n,m}}{2} \left ( \prod_{r \in S_L} 2 \cosh \left ( \frac{\beta \epsilon_{q_r}}{2}\right ) \frac{1}{1+e^{\beta \epsilon_{q_n}}} -  \prod_{r \in S_L} 2 \sinh \left ( \frac{\beta \epsilon_{q_r}}{2}\right ) \frac{1}{1-e^{\beta \epsilon_{q_n}}}\right ).
    \end{aligned}
    \label{eq:SM:expectation-value-in-one-parity-sector}
    \end{equation}
The last ingredient is the modified partition function, which can be expressed as
    \begin{equation}
    \begin{aligned}
        &\mathrm{Tr} \left (e^{-\beta H} e^{i \pi N} \right ) = \mathrm{Tr} \left (P_\text{e} e^{-\beta H_\text{e}} \right )-\mathrm{Tr} \left (P_\text{o}e^{-\beta H_\text{o}}\right ) = \\ &= \frac{1}{2} \left ( \prod_{r \in S_L} 2 \cosh \left ( \frac{\beta \epsilon_{p_r}}{2}\right )  +  \prod_{r \in S_L} 2 \sinh \left ( \frac{\beta \epsilon_{p_r}}{2}\right ) \right ) - \frac{1}{2} \left ( \prod_{r \in S_L} 2 \cosh \left ( \frac{\beta \epsilon_{q_r}}{2}\right )  -  \prod_{r \in S_L} 2 \sinh \left ( \frac{\beta \epsilon_{q_r}}{2}\right ) \right ).
    \end{aligned}
    \label{eq:SM:modified-partition-function}
    \end{equation}

Combining Eqs.\ (\ref{eq:SM:fermionic-part-derivation-free-fermion-formula}), (\ref{eq:SM:time-evolution-approx}), (\ref{eq:SM:modified-thermal-exp-value-of-general-operator}), (\ref{eq:SM:expectation-value-in-one-parity-sector}) and (\ref{eq:SM:modified-partition-function}), we can calculate the $L \rightarrow \infty$ limit of Eq.\ (\ref{eq:SM:fatorization-hypothesis-fermionic-part}). A fully analytical treatment is beyond the scope of this work, but high-precision numerics (based on the implementation of these formulae) show that as $L \rightarrow \infty$,
    \begin{equation}
        F_\text{d}(x,t) \rightarrow -\int_{-\pi}^\pi \frac{dp}{2\pi} \left [ \frac{e^{i \theta_p} e^{ipx-i \epsilon(p)t}}{1-e^{\beta \epsilon(p)}} +\frac{-e^{i \theta_{p}} e^{ipx+i \epsilon(p)t}}{1-e^{-\beta \epsilon(p)}}\right],\;\;\text{for}\;\; h>1,
    \label{eq:SM:fermionic-part-result-for-paramagnet}
    \end{equation}
which is what we would get by a naive calculation, disregarding the presence of the two sectors, but 
    \begin{equation}
        F_\text{o}(x,t) \rightarrow e^{x \delta}\;\;\text{for}\;\; h<1,
    \label{eq:SM:fermionic-part-result-for-ferromagnet}
    \end{equation}
where $\delta$ is defined in Eq.\ (\ref{eq:SM:ferromagnetic-degenerate-state}). This means that for $h < 1$, Eq.\ (\ref{eq:SM:fatorization-hypothesis-fermionic-part}) is governed by the effects of the zero-mode $\ket{\Delta}$, and the other part of the spectrum is completely irrelevant. 

Combining Eqs.\ (\ref{eq:SM:factorization-hypothesis}), (\ref{eq:SM:string-part-for-paramagnet}), (\ref{eq:SM:string-part-for-ferromagnet}), (\ref{eq:SM:fermionic-part-result-for-paramagnet}) and (\ref{eq:SM:fermionic-part-result-for-ferromagnet}), we get that in the hydrodynamic limit,
    \begin{equation}
    \begin{aligned}
        \braket{\sigma_1^1 \sigma_{1+x}^1(t)} &\asymp F_\text{d}(x,t) \Omega_\text{d}(x,t),\;\;\;\qquad\qquad\text{for}\;\;h>1,\\
        \braket{\sigma_1^1 \sigma_{1+x}^1(t)} &\asymp F_\text{o}(x,t) \Omega_\text{o}(x,t) = \Omega_\text{d}(x,t),\;\;\text{for}\;\;h<1,
    \end{aligned}
    \label{eq:SM:factorization-end-result}
    \end{equation}
which shows that the exponential contributions of the ferromagnetic zero-mode cancel.

\section{Evaluating the correlation length}

The asymptotic behavior of Eq.\ (\ref{eq:SM:factorization-hypothesis}) is an exponential with a correlation length that depends on the space-time ray $\zeta = t/x$, the magnetic field, and the temperature:
    \begin{equation}
        \braket{\sigma_a^{1} \sigma_{1+x}^1(t = \zeta x)} \asymp \exp \left ( -x/\xi(\zeta, h, \beta)\right ).
    \label{eq:SM:corr-length-def}
    \end{equation}
We would like to extract the behavior of the correlation length as a function of its three arguments. 

Using Eq.\ (\ref{eq:SM:factorization-end-result}), we need to calculate the asymptotic behavior of $\Omega_\text{d}(x,t = \zeta x)$ and $F_\text{d}(x, t = \zeta x)$ (for $h > 1$). The fluctuation part is less interesting; it is given by the integral
    \begin{equation}
        \xi_\text{fl}^{-1}(\zeta, h, \beta) = -\int_{-\pi}^\pi \frac{dq}{2\pi} |1-\epsilon'(q)| \log \tanh \left ( \frac{\beta \epsilon(q)}{2} \right ).
    \label{eq:SM:BFT-corr-length-def}
    \end{equation}
The propagation contribution 
    \begin{equation}
        F_\text{d}(x, t = \zeta x) = \exp \left ( -x/\xi_\text{pr}(\zeta, h, \beta) \right )
    \label{eq:SM:fermionic-corr-length-def}
    \end{equation}
is much more complicated, so we dedicate the rest of this section to its asymptotic analysis.

First, we label the two contributions of Eq.\ (\ref{eq:SM:fermionic-part-result-for-paramagnet}) as
    \begin{equation}
        F_\text{d}(x, t = \zeta x)  = -\int_{-\pi}^\pi \frac{dp}{2\pi} \left [ \frac{e^{i \theta_p} e^{ipx-i \epsilon(p)t}}{1-e^{\beta \epsilon(p)}} +\frac{-e^{i \theta_{p}} e^{ipx+i \epsilon(p)t}}{1-e^{-\beta \epsilon(p)}}\right] := I^{(1)} + I^{(2)}
    \label{eq:SM:fermionic-integra-two-contributions-def}
    \end{equation}
and notice that $I^{(1)}$ and $I^{(2)}$ are connected by an $\epsilon(p) \rightarrow -\epsilon(p)$ transformation. (The extra minus sign appears because Eq.\ (\ref{eq:SM:Bogoliubov-angle-cos}) contains $\epsilon(p)$ in the denominator.) This is equivalent to changing between the complex sheets generated by the square root in the dispersion relation. Therefore, Eq.\ (\ref{eq:SM:fermionic-integra-two-contributions-def}) can be evaluated by integrating 
    \begin{equation}
        f(z) = \frac{1}{2\pi} \frac{e^{i \theta_z} e^{izx-i \epsilon(z)\zeta x}}{1-e^{\beta \epsilon(z)}}
        \label{eq:SM-function-to-be-complex-intergated}
    \end{equation}
between $-\pi$ and $\pi$ on both sheets and summing the results.

We use the method of steepest descent to extract the asymptotic decay of Eq.\ (\ref{eq:SM:fermionic-integra-two-contributions-def}) as $x \rightarrow \infty$. First, we define the exponents in the two sheets as
    \begin{equation}
        \Phi^{(1,2)}(z) = i z \mp i \epsilon(z) \zeta,
    \label{eq:SM:phase-in-the-two-sheets}
    \end{equation}
where $\epsilon(z)$ represents the complex function evaluated on the first sheet and the -(+) sign corresponds to the first (second) sheet. Then, $\partial_z\Phi^{(1,2)}(z) = 0$ gives the two saddle points situated in positions $z_{S1}$ and $z_{S2}$ satisfying
    \begin{equation}
    \begin{aligned}   
        \cos z_{S1} &= \frac{1+\sqrt{(1-\zeta^2)(1-\zeta^2 h^2)}}{h \zeta^2} ,\\
            \cos z_{S2} &= \frac{1-\sqrt{(1-\zeta^2)(1-\zeta^2 h^2)}}{h \zeta^2}.
    \end{aligned}
    \label{eq:SM:saddle-point-positions}
    \end{equation}
We choose the solutions satisfying $\mathrm{Re}\,  z_{S1},\,\mathrm{Re}\, z_{S2} \in [-\pi,\,\pi]$ and $\mathrm{Im}\,  z_{S1},\,\mathrm{Im}\, z_{S2}\geq 0$. Notice that their position is identical on both sheets. 

There are three distinct parameter regimes to consider.
    \begin{itemize}
        \item If $1/\zeta<1$, both saddle points have zero imaginary parts and are within the integration domains: $z_{S1}, \,z_{S2} \in [-\pi,\, \pi]$. Then, the method of stationary phase implies that the decay of Eq.\ (\ref{eq:SM:fermionic-integra-two-contributions-def}) will be algebraic in $x$, so it will not contribute to the correlation length $\xi_\text{pr}$.
        \item For $1<1/\zeta<h$, the saddle points acquire a nonzero imaginary part, and the $\Phi^{(1,2)}$ exponents evaluated at these points give
            \begin{equation}
            \begin{aligned}
               \text{Re}\,  \Phi^{(1,2)}(z_{S1}) &= \mp \sqrt{1-\zeta^2}-\mathrm{arccosh}(1/\zeta),\\
               \text{Re}\,  \Phi^{(1,2)}(z_{S2}) &= \pm \sqrt{1-\zeta^2}-\mathrm{arccosh}(1/\zeta),\\
               \text{Im}\,  \Phi^{(1,2)}(z_{S1}) &= -\arccos (1/(h \zeta)) \mp \sqrt{\left (h \zeta\right )^2 -1},\\
               \text{Im}\,  \Phi^{(1,2)}(z_{S2}) &= \arccos (1/(h \zeta)) \mp \sqrt{\left (h \zeta\right )^2 -1}.
            \end{aligned}
            \label{eq:SM:saddle-point-phi-values-large-h}
            \end{equation}
        As we will see, the decay of Eq.\ (\ref{eq:SM:fermionic-integra-two-contributions-def}) will be dominated by $S1$ in the first sheet and by $S2$ in the second sheet, and hence will be exponential in $x$, giving a finite contribution to the correlation length.
        \item If $1<h<1/\zeta$, the saddle points become purely imaginary satisfying
            \begin{equation}
                0<\text{Im}\, z_{S2}<\text{Im}\, z_{S1},\qquad \text{Re} \,z_{S1} = \text{Re}\, z_{S_2} = 0. 
            \label{eq:SM:saddle-point-position-middle-h}
            \end{equation}
        Evaluating the $\Phi^{(1,2)}$ functions at these points, we get
            \begin{equation}
            \begin{aligned}
                \Phi^{(1,2)}(z_{S1}) &= \mp w_1(\zeta,h) \pm \log \left [\frac{1+\sqrt{(1-\zeta^2)(1-(\zeta h)^2)}\mp w_1(\zeta,h)}{h \zeta^2}\right] \in \mathbb{R},\\
                \Phi^{(1,2)}(z_{S2}) &= \pm w_2(\zeta,h) \mp \log \left [\frac{1-\sqrt{(1-\zeta^2)(1-(\zeta h)^2)}\pm w_1(\zeta,h)}{h \zeta^2}\right] \in \mathbb{R},
            \end{aligned}
            \label{eq:sM:saddle-point-phi-values-medium-h}
            \end{equation}
        where 
            \begin{equation}
            \begin{aligned}
                w_1(\zeta,h) = \sqrt{2-(1+h^2)\zeta^2 +2\sqrt{(1-\zeta^2)(1-(\zeta h)^2)}},\\
                w_2(\zeta,h) = \sqrt{2-(1+h^2)\zeta^2 -2\sqrt{(1-\zeta^2)(1-(\zeta h)^2)}}.
                \label{eq:SM:omega-fucntions-def}
            \end{aligned}
            \end{equation}
        Once again, the decay of Eq.\ (\ref{eq:SM:fermionic-integra-two-contributions-def}) will be exponential, and there will be an interesting competition between the saddle points and the poles of Eq.\ (\ref{eq:SM-function-to-be-complex-intergated}) to give the leading contribution.
    \end{itemize}

Next, we have to discuss the pole structure of Eq.\ (\ref{eq:SM-function-to-be-complex-intergated}). It is easy to see that poles correspond to solutions of the 
    \begin{equation}
    e^{\beta \epsilon(z)} = 1
    \label{eq:SM:pole-condition}
    \end{equation}
equation, identically on both sheets. We only consider solutions with real parts between $-\pi$ and $\pi$ and positive imaginary parts. Labeling the $n$th solution with $z_{Pn}$, we can write
    \begin{equation}
        z_{Pn} = \arccos \left ( \frac{(1+h^2)\beta^2  + 4 \pi^2 n^2}{2h\beta^2}\right ) \in i \mathbb{R}, \qquad (n \geq 0, n \in \mathbb{N}).
    \label{eq:SM:pole-positions}
    \end{equation}
This implies that these poles lie in the imaginary axis and their imaginary part increases with $n$. Furthermore, the $\Phi^{(1,2)}$ exponents evaluated at these poles give
    \begin{equation}
        \Phi^{(1,2)}(z_{Pn}) = \pm\frac{2\pi n \zeta}{\beta} - \mathrm{arccosh} \left ( \frac{(1+h^2)\beta^2  + 4 \pi^2 n^2}{2h\beta^2}\right ) \in \mathbb{R}.
    \label{eq:SM:pole-exponent-values}
    \end{equation}
Importantly, the branch cut of Eq.\ (\ref{eq:SM-function-to-be-complex-intergated}) starts at $z_{P0}$ and runs along the imaginary axis. 

Finally, to apply the method of steepest descent, we have to consider the constant phase contours satisfying
    \begin{equation}
        \mathrm{Im} \,\Phi^{(1,2)} = \text{const}.
    \label{eq:SM:constant-phase-contour-def}
    \end{equation}
To evaluate Eq.\ (\ref{eq:SM:fermionic-integra-two-contributions-def}), we need to find such contours on both sheets that connect $-\pi$ and $\pi$.

        \begin{figure}[t]
          \centering
          \subfloat[Contour for $I^{(1)}$.]{
            \includegraphics[width=0.45\linewidth]{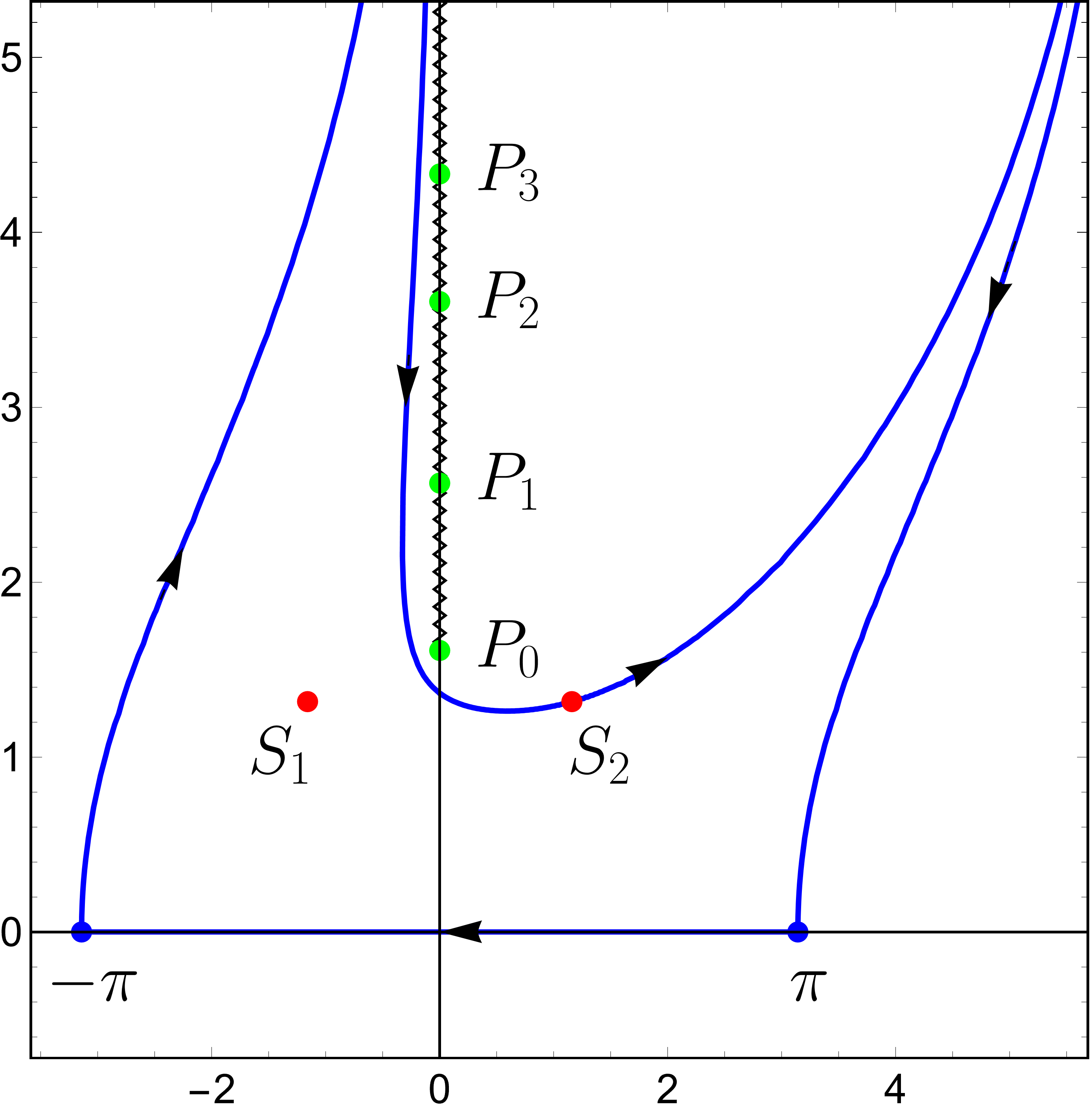}
            \label{images:largehsheet1}
          }
          \hfill
          \subfloat[Contour for $I^{(2)}$.]{
            \includegraphics[width=0.45\linewidth]{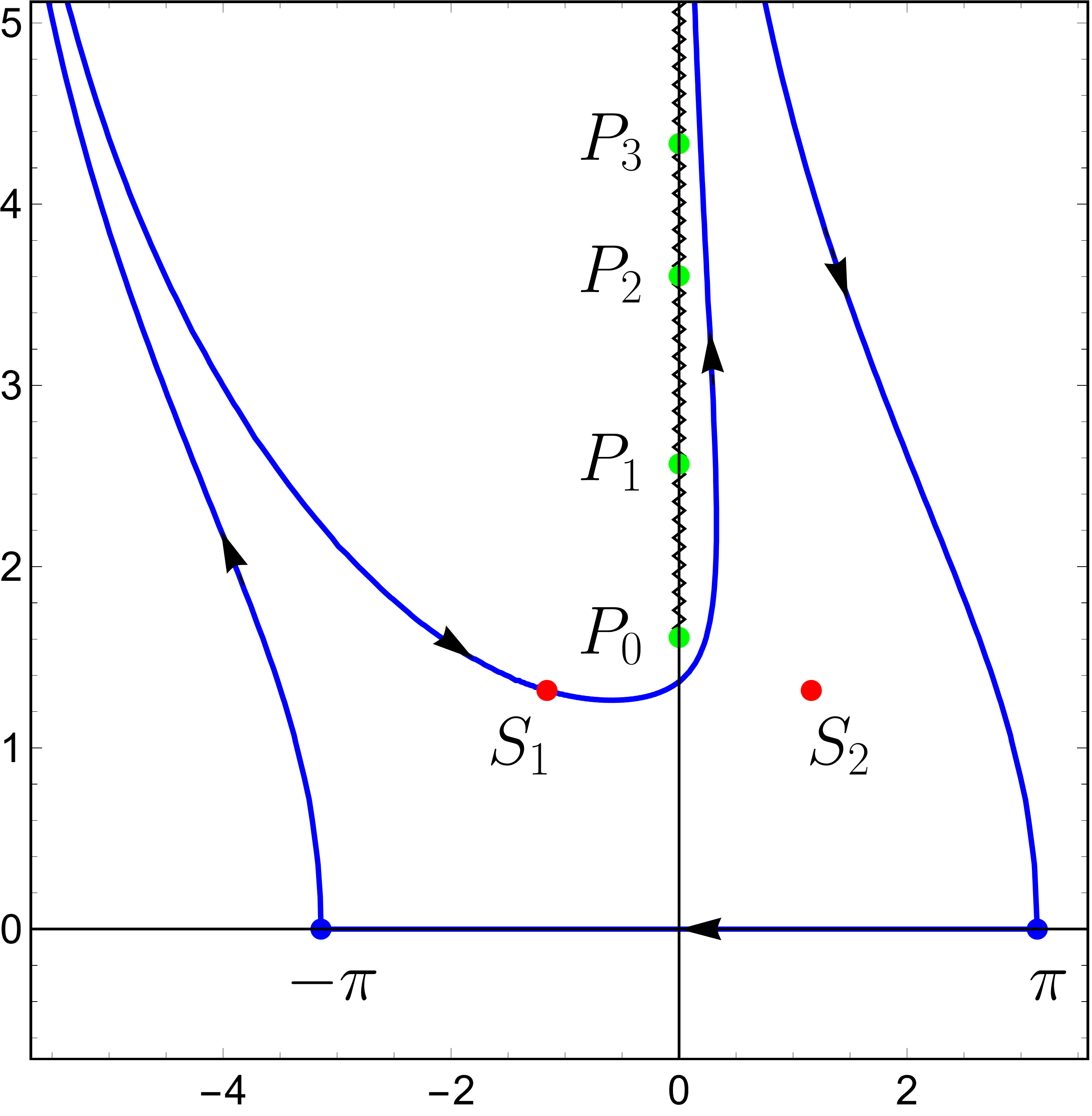}
            \label{images:largehsheet2}
          }
          \caption{Integration contours for evaluating Eq.~(\ref{eq:SM:fermionic-integra-two-contributions-def}) for $1<1/\zeta<h$ using the method of steepest descent. The blue lines correspond to the constant phase contours. $I^{(1)}$ and $I^{(2)}$ are dominated by the saddle points $S_2$ and $S_1$. The contours avoid the branch cut, so they can be closed in the two sheets separately.}
          \label{fig:SM:large-h-contours}
        \end{figure}

First, we consider the case $1<1/\zeta<h$, where the saddle points have non-zero real and imaginary parts. Figure \ref{fig:SM:large-h-contours} depicts the integration contours, the saddle points, the poles, and the branch cut. As we can see, these contours avoid the branch cut, so they can be closed separately on the two sheets, allowing the asymptotic calculation of both $I^{(1)}$ and $I^{(2)}$ separately. Furthermore, the contours do not encircle any poles, so Cauchy's theorem ensures that the result of the integrals vanish on both sheets. Therefore, $I^{(1,2)}$ equals minus the contributions of the constant phase contours on the respective sheet. However, the method of steepest descent implies that the leading behavior of these integrals is dominated by $S2$ in the first and $S1$ in the second sheets. (The contributions of the paths starting from $\pm \pi$ cancel due to the $2\pi$ periodicity of the integrands.) Using Eq.\ (\ref{eq:SM:saddle-point-phi-values-large-h}), we get
    \begin{equation}
    \begin{aligned}
        I^{(1)} \asymp \frac{C^{(1)}}{\sqrt{x}} e^{x\left ( \sqrt{1-\zeta^2} -\text{arccosh}(1/\zeta) \right )}e^{i x \left ( \arccos (1/(h \zeta)) - \sqrt{\left (h \zeta\right )^2 -1} \right )},\\
        I^{(2)} \asymp \frac{C^{(2)}}{\sqrt{x}}  e^{x\left ( \sqrt{1-\zeta^2} -\text{arccosh}(1/\zeta) \right )}e^{-i x \left ( \arccos (1/(h \zeta)) - \sqrt{\left (h \zeta\right )^2 -1} \right )},
    \label{eq:SM:result-for-large-h}
    \end{aligned}
    \end{equation}
where $C^{(1,2)}$ are constants coming from the Gauss integrals in the vicinity of the saddle points. Note that the inverse correlation length for both $I^{(1)}$ and $I^{(2)}$ are the same,
    \begin{equation}
        \xi_\text{pr}^{-1}(\zeta, h, \beta)\bigg |_{1 < 1/\zeta < h} = -\sqrt{1-\zeta^2} + \text{arccosh}(1/\zeta)>0,
    \label{eq:SM:corr-length-contribution-large-h}
    \end{equation}
but the oscillatory terms differ in the two sheets. Interestingly, the correlation length is independent of both $\beta$ and $h$. In summary, the leading decay of Eq.\ (\ref{eq:SM:fermionic-integra-two-contributions-def}) for $1<1/\zeta < h$ can be written as
    \begin{equation}
        F_\text{d}(x,t = \zeta x)\bigg |_{1<1/\zeta < h} \asymp \frac{e^{-x/\xi_\text{pr}(\zeta)} }{\sqrt{x}} \left ( C^{(1)} e^{i x (\arccos (1/(h\zeta))-\sqrt{(h \zeta)^2-1})}+C^{(2)}e^{-i x (\arccos (1/(h\zeta))-\sqrt{(h \zeta)^2-1})}\right ).
    \label{eq:SM:large-h-leading-term-sum}
    \end{equation}
Note that only the $C^{(1,2)}$ constants have any temperature dependence.

\begin{figure}[t]
      \centering
      \subfloat[Contour on the first sheet.]{
        \includegraphics[width=0.45\linewidth]{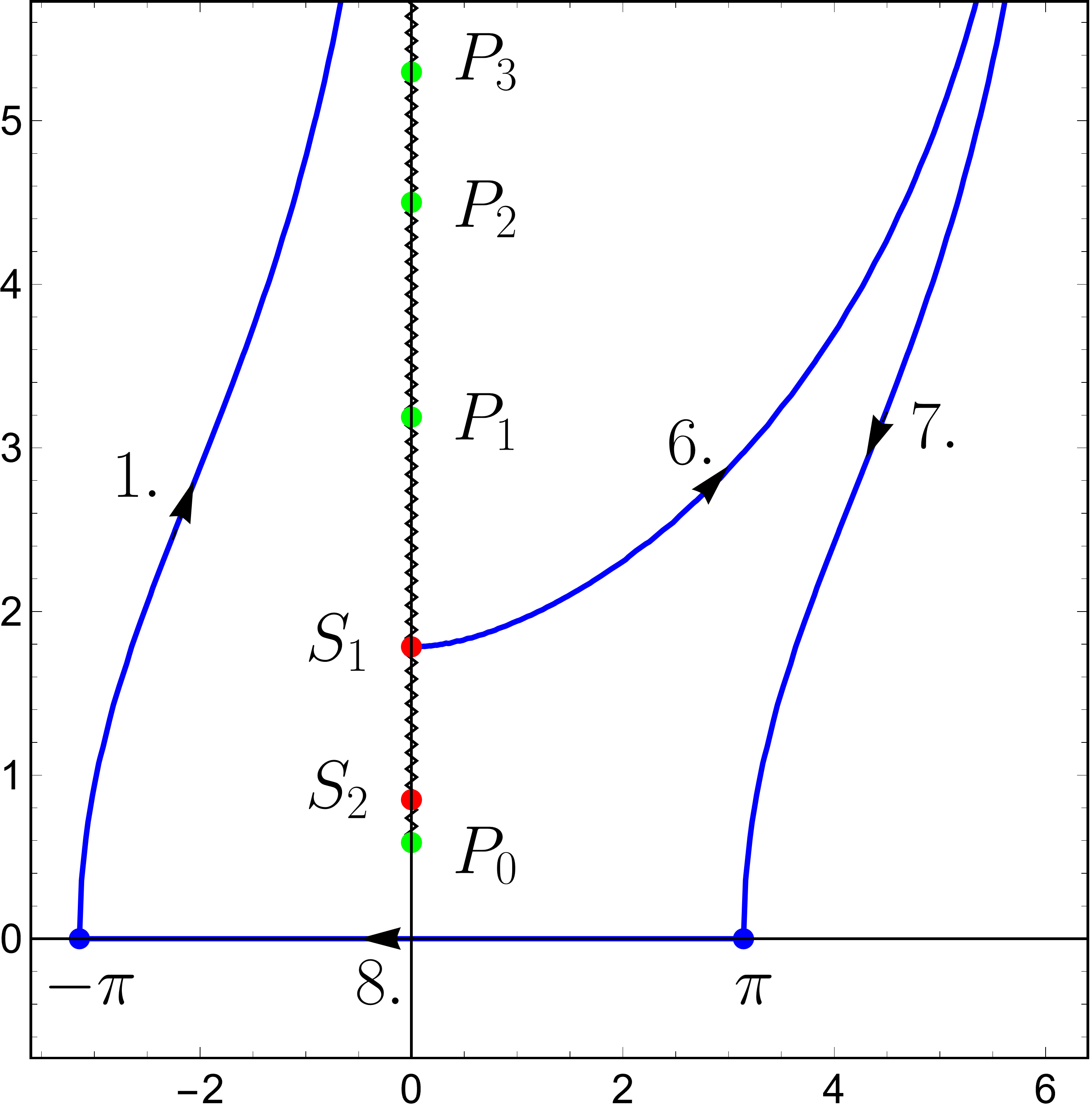}
        \label{images:mediumhsheet1}
      }
      \hfill
      \subfloat[Contour on the second sheet.]{
        \includegraphics[width=0.45\linewidth]{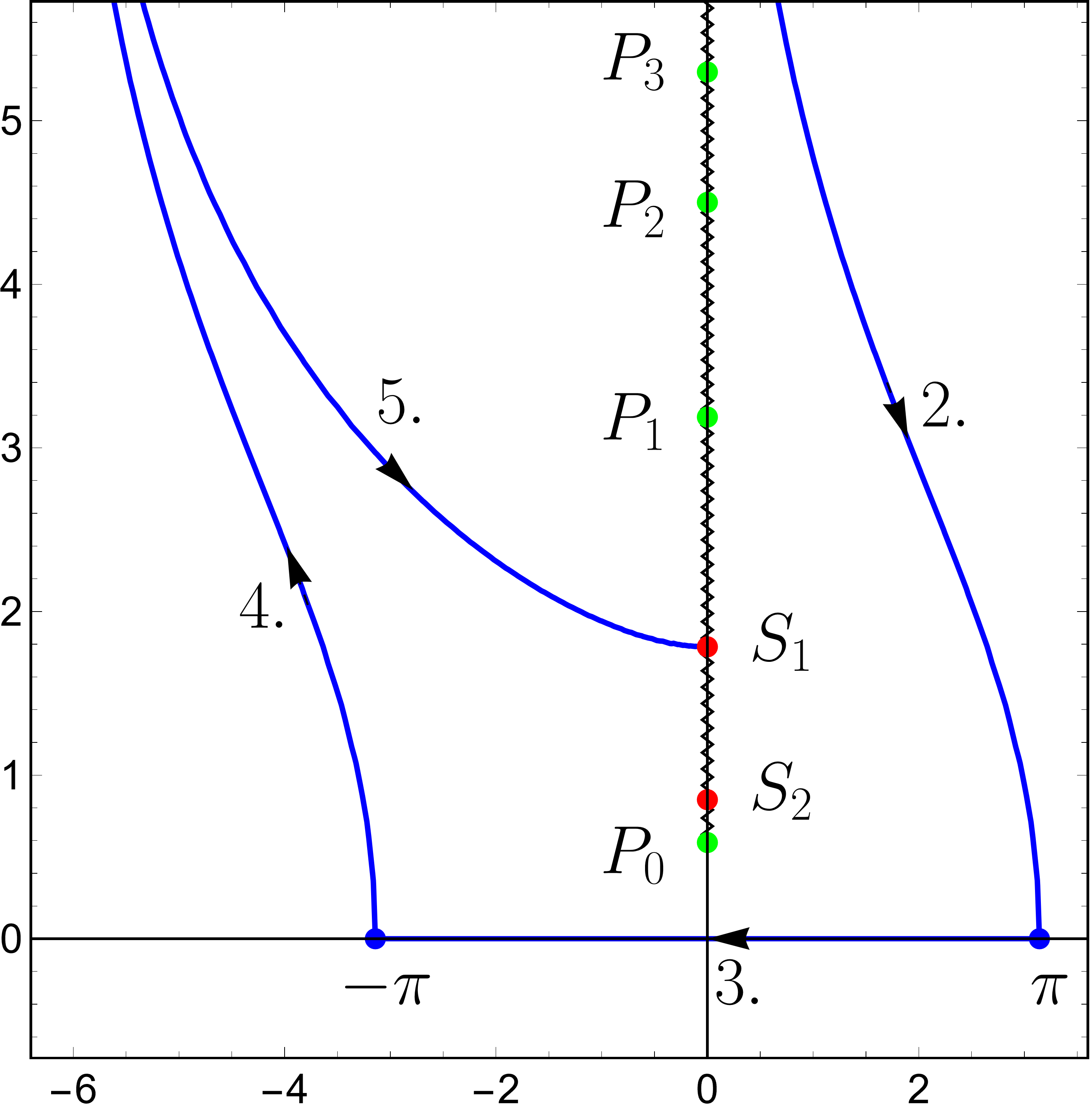}
        \label{images:mediumhsheet2}
      }
      \caption{Integration contours for evaluating Eq.~(\ref{eq:SM:fermionic-integra-two-contributions-def}) for $1<h<1/\zeta$ using the method of steepest descent.}
      \label{fig:SM:medium-h-contours}
    \end{figure}

Lastly, we consider the case $1<h<1/\zeta$. Here, the constant phase contours are more complicated because they cross the branch cut as depicted in Figure \ref{fig:SM:medium-h-contours}. Therefore, the integration contour can be closed only on the Riemann surface as a whole, not on the separate sheets. Furthermore, the contour will encircle some poles that compete for the leading contribution.

By the continuity of the integrands, paths 1 and 2 encircle all poles on the second sheet, but the poles of the first sheet contribute only if they lie below $S1$. We can count these with the function
    \begin{equation}
        n_{\text{max}} = \max \left \{n\; |\; n \in \mathbb{N},\; \text{Im}\,z_{Pn}<\text{Im}\,z_{S1} \right \}.
    \label{eq:SM:nmax-intro}
    \end{equation}
For example in Figure \ref{fig:SM:medium-h-contours} $n_\text{max}=0.$ 

Furthermore, the continuity of Eq.\ (\ref{eq:SM-function-to-be-complex-intergated}) on the Riemann surface implies that the contour only crosses $S1$ on the first sheet and not on the second one. Therefore, we have to find the maximum value among the real numbers $\{\Phi^{(2)}(z_{Pn}) | n \in \mathbb{N}\}$, $\{\Phi^{(1)}(z_{P0}),\dots,\Phi^{(1)}(z_{Pn_\text{max}})\}$, and $\Phi^{(1)}(z_{S1})$. However, Eq.\ (\ref{eq:SM:pole-exponent-values}) implies that for all $n,k>0$, $\Phi^{(2)}(z_{Pn})<\Phi^{(1)}(z_{Pk})$, which means that the second sheet can be completely excluded from the asymptotic analysis. Then the inverse correlation length is given by
    \begin{equation}
    \xi_\text{pr}^{-1}(\zeta, h, \beta)\bigg|_{1<h<1/\zeta} = \max \{\Phi^{(1)}(z_{P0}),\,\dots,\Phi^{(1)}(z_{Pn_\text{max}}),\,\Phi^{(1)}(z_{S1})\},
    \label{eq:SM:middle-h-corr-length}
    \end{equation}
and the leading decay of Eq.\ (\ref{eq:SM:fermionic-integra-two-contributions-def}) is given by
    \begin{equation}
        F_\text{d}(x,t= \zeta x)\bigg |_{1<h<1/\zeta} \asymp C e^{-x/\xi_\text{pr}(\zeta,h,\beta)},
    \label{eq:SM:middle-h-leading-term}
    \end{equation}
where $C$ is an undetermined prefactor.

\end{document}